\documentclass[
fleqn
]{article}

\usepackage{mymacros,myunicode}
\usepackage{graphicx,color}

\renewenvironment{mybox}{\begin{array}{|l|}\hline} {\\\hline\end{array}}
\newenvironment{multiequation}{\begin{equation}\begin{array}{l}}{\end{array}\end{equation}}

\newcommand{\tsr}{\mathsf}   
\newcommand{\vt}{\boldsymbol} 
\newcommand{\ve}{\vt{\mathrm{e}}}
\newcommand{\be}{\begin{equation}}
\newcommand{\ee}{\end{equation}}

\newcommand{\m}{\rule[3pt]{.3em}{.5pt}}

\renewenvironment{eqnarray}{  \begin{equation}\begin{array}{l}  }{  \end{array}\end{equation}  }

\renewcommand{\u}{\vt{u}} 
\newcommand{\x}{\vt{x}}

\renewcommand{\hat}{\widehat}
\newcommand{\pdfrac}[2] {\frac{\partial #1}{\partial #2}\,}

\title{Group theoretic approach and analytical solutions of the compressible Navier--Stokes equations}
\author{Dina Razafindralandy, Aziz Hamdouni\\
 La Rochelle Université -- France}
\date{}

\begin{document}

\maketitle

\begin{abstract}
	A group theoretic analysis of the compressible Navier-Stokes equations of an ideal gas are carried out. The 12-dimensional Lie symmetry group is computed. The commutation table and the Levi decomposition of its Lie algebra are presented. The equations are reduced and self-similar one-, two- and three-dimensional solutions are computed. Many of them are graphically illustrated.

\end{abstract}


\section{Introduction}

Fluid flows have a important role in engineering science. To understand the complex phenomena that arise in them, computing analytical solutions may be helpful. Indeed, an analytical solution may serve as a simplified model of the flow in a particular configuration. It may also be used to calibrate or tests numerical schemes or turbulent models.

Some methods exist in finding particular solutions of partial differential equations (separation of variables, assuming axisymmetry, ...). Most of them consist in choosing an ansatz and reduce the equations accordingly. However, finding ansätze which effectively reduce the equation is not an easy task. A tool which provides a systematic way of finding the such ansätze is the Lie symmetry-group theory \cite{olver86,ibragimov}. 

In incompressible fluid mechanics, the Lie group theory has been used to find analytical solutions of the Navier--Stokes equations \cite{fushchych94}, and particularly vortex-like ones \cite{grassi00}, to study turbulence \cite{unal97,oberlack01,ejm07} or to model particular flows such as boundary layers \cite{khujadze04} and thin isothermal and non-isothermal shear layers \cite{nova09}. In this article, we deal with the compressible case where only few exact solutions are available. The reader may refer to some papers such as \cite{colonius91,curle71,tsangaris00,sachdev05} to find analytical solutions.  Most of them are either one-dimensional or axisymmetric. The aim of the present work is to carry out a group theoretic analysis of the governing equations and to propose a wide class of analytical solutions.

In section \ref{section:equation}, the hypotheses on the flow are listed and the velocity-pressure-density formulation of the compressible Navier-Stokes equations is presented. In section \ref{symmetry}, the Lie method of symmetry computation is briefly recalled. The Lie point-symmetry group of the equations and its Lie algebra are then analysed. It will be shown that the Lie-algebra is solvable. The symmetry group will be used to find ansätze and reduce the equations in the subsequent sections. We do not intend to be exhaustive. Rather, some interesting self-similar solutions, such as vortex-like ones, are presented, in order to complete the set of available solutions in the literature. In sections \ref{steady}, \ref{unsteady} and \ref{3d}, respectively steady bidimensional, unsteady bidimensional and three-dimensional cases are considered.

\section{Model equations\label{section:equation}}


The motion of a fluid is governed by the Navier-Stokes equations \cite{batchelor_2000}: 
\begin{equation}\begin{cases}
\td{\rho}{t}+\rho\div\u=0 
\\\\
\rho\td{\u}{t}= \div σ
\\\\\displaystyle 
ρ\td {}t⦅e+÷{\u^2}2⦆=\div(σ\u+κ∇ T)
\end{cases}\label{comp0}
\end{equation}
in absence of body force. In these equations, $\rho$, $\u=(u,v,w)$, $e$ and $T$ are respectively the density, the velocity, the internal energy and the temperature. $κ$ is the thermal diffusion coefficient, taken constant.
With the hypothesis of a Newtonian fluid, the stress tensor $\sigma$ writes:
\begin{equation}
 \sigma=\left(-p-\cfrac{2\mu}{3}\,\div\u\right)\tsr{I_d}+2\mu\tsr{S}.
\end{equation}
where $p$ is the pressure field, µ is the dynamic viscosity, $\tsr{I_d}$ is the three-dimensional identity matrix and $\tsr{S}$ is the strain rate tensor
\begin{equation}
	\tsr{S}=\cfrac{\nabla\vt{u}+\tpleft{\,\nabla\vt{u}}}{2}.
\end{equation}
The variation of $µ$ with the temperature is neglected. Equations (\ref{comp}) can also be formulated as follows
\begin{equation}\begin{cases}
\td{\rho}{t}+\rho\div\u=0 
\\\\
\rho\td{\u}{t}= \div σ
\\\\\displaystyle 
ρ\td et=σ:\tsr S+κΔT
\end{cases}\label{comp}
\end{equation}
where the double dot sign stands for the Frobenius inner product:
\[ \sigma:\tsr S=\operatorname{tr}(\tpleft{\sigma}\ \tsr S)=∑_{i,j}\sigma_{ij}S_{ij}. \]
Assume that the fluid is an ideal gas. We then have:
\begin{equation}
	p=ρTR \label{p}
\end{equation}
where $R$ is the gas constant. Moreover, the internal energy is proportional to the temperature, that is
\begin{equation}
	e=C_vT \label{e}
\end{equation}
where $C_v$ is the (constant) specific heat at constant volume. 
Inserting relations (\ref{e}) and (\ref{p}) in the energy equation of (\ref{comp}) and using the mass balance equation, we get the density-velocity-pressure formulation (see also \cite{toutant17,zhang92,garnier09}):
\begin{equation}\begin{cases}
\td{\rho}{t}+\rho\div\u=0 
\\\\
\rho\td{\u}{t}= \div σ
\\\\
\cfrac{C_v}{R}\left(\td{p}{t}+p\div \u\right)=\sigma:\tsr{S}+\cfrac{\kappa}{R}\ \Delta \left(\cfrac{p}{\rho}\right)
\end{cases}\label{nsc}
\end{equation}
Expliciting material derivatives with the relation
\begin{equation}
	\td{}t=\pd {}t+(\u\cdot ∇),
\end{equation} one gets a (five-dimensional) partial differential equation
\begin{equation}
	\mathbf E⦅t,\x,\u_{(2)},p_{(2)},ρ_{(2)}⦆=0
	\label{nsc3}
\end{equation}
where $\u_{(2)}$, $p_{(2)}$ and $ρ_{(2)}$ gather $u$, $v$, $w$, $p$, $ρ$ and all of their partial derivatives up to second order. A componentwise expression of equation (\ref{nsc3}) is given in appendix \ref{componentwise}, equation (\ref{nsc2}).

\section{Lie symmetry group and Lie algebra\label{symmetry}}

In this section, we study the Lie symmetry group admitted by equations (\ref{nsc}) and its Lie algebra. The suited framework to compute Lie symmetry groups is the language of jet space but to simplify the presentation, we avoid its introduction. More details can be found in \cite{olver86,bluman02,hydon00a,ibragimov}.

A transformation
\begin{equation}
 T\ :\ (t,\x,\u,p,\rho)\longmapsto(\hat{t},\hat{\x},\hat{\u},\hat{p},\hat{\rho})
\label{transformation}\end{equation}
is called a (point-)symmetry of equations (\ref{nsc3}) if it transforms any solution of  (\ref{nsc3}) into another solution, that is 
\begin{equation}
	\mathbf E (t,\x,\u_{(2)},p_{(2)},\rho_{(2)})=0\quad\quad\Longrightarrow\quad\quad \mathbf E(\hat{t},\hat{\x},\hat{\u_{(2)}},\hat{p_{(2)}},\hat{\rho_{(2)}})=0.\label{sym}
\end{equation}
where $\hat{\u_{(2)}}$, $\hat{p_{(2)}}$ and $\hat{\rho_{(2)}}$ represent the transforms $\hat{\u}$, $\hat{p}$, $\hat{\rho}$ of $\u$, $p$, $ρ$ and their partial derivatives with respect to $\hat t$ and $\hat{\x}$ up to second order. Our aim is to find all the local Lie symmetry groups of (\ref{nsc}), that are families of symmetries
\[ G=\{T_{ϵ} : (t,\x,\u,p,\rho)\longmapsto(\hat{t},\hat{\x},\hat{\u},\hat{p},\hat{\rho})\ |\ ϵ∈I⊂ℝ,\, T_{ϵ} \text{ symmetry of (\ref{nsc})}\}\]
having a structure of a local Lie group. For the sake of simplicity, we assume that the group is additive. In particular, $0∈I$ and $T_{ϵ=0}$ is the identity transformation.

Computing local Lie symmetrie group is generally easier if symmetry condition (\ref{sym}) is linearized at the vicinity of the identity. To this aim, consider the vector field
\begin{equation}X=\xi^t\pd{}{t}+\xi^x\pd{}{x}+\xi^y\pd{}{y}+\xi^z\pd{}{z} +\xi_u\pd{}{u}+\xi_v\pd{}{v}+\xi_w\pd{}{w} +\xi_p\pd{}{p}+\xi_\rho\pd{}{\rho} \label{generator}\end{equation}
where the components
\be 
\xi^r(t,\x,\u,p,ρ)=\left.\td{\hat{r}}{ϵ}\right|_{ϵ=0} ,\quad r=t,x,y,z
\ee
represent the infinitesimal variation of the independant variables under the action of $G$ and
\be 
\xi_q(t,\x,\u,p,ρ)=\left.\td{\hat{q}}{ϵ}\right|_{ϵ=0} ,\quad q=u,v,w,p,ρ
\ee
represent those of the dependant variables.  Vector field $X$ is called the generator of $G$. According to the Lie group theory \cite{olver86,ibragimov}, if $G$ is a Lie symmetry of (\ref{nsc}) then
\begin{equation}
  \left.pr^{(2)}X\cdot \mathbf E\right|_{\mathbf E=0}=0
\label{symcond}\end{equation}
where $pr^{(2)}X$ is the second-order prolongation of $X$. It writes:
\[ pr^{(2)}X=X+X^{(1)}+X^{(2)} \]
where
\begin{equation}
	X^{(1)}=∑_q∑_{s}ξ^s_q\pdfrac{}{q_s}
	\label{x1}
\end{equation}
takes into account the infinitesimal variation of first order partial derivatives under the action of $G$  (see also equation (\ref{x1components})) and
\begin{equation}
	X^{(2)}=\displaystyle∑_q∑_{r,s}\xi^{rs}_q\pdfrac{}{q_{rs}}
	\label{x2}
\end{equation}
acts on second order derivatives. 
In (\ref{x1}) and (\ref{x2}), the sums are over all dependent variables $q=u,v,w,p,ρ$ and over all independent variables $r,s=t,x,y,z$.
The coefficients of $X^{(1)}$ and $X^{(2)}$ are linked to those of $X$ by the relations:
\be \xi_q^s=D_s\xi_q-\sum_{m=t,x,y,z}q_mD_s\xi^m,\label{x1relation}\ee
\be \xi_q^{rs}=D_s\xi_q^r-\sum_{m=t,x,y,z}q_{rm}D_s\xi^m, \label{x2relation}\ee
$D_r$ being the total derivation operator with respect to $r$.

Infinitesimal symmetry condition (\ref{symcond}) applied to (\ref{nsc}) leads to system of partial differential equations on the $\xi^r$ and $ξ_q$. After solving this system, one finds that $X$ generates a Lie symmetry of (\ref{nsc}) if it is a linear combination of the following vector fields (see appendix \ref{componentwise}):
\be X_1=\pd{}{t}, \qquad X_2=\pd{}{x}, \qquad X_3=\pd{}{y}, \qquad X_4=\pd{}{z}, \nonumber\ee
\be X_5=t\pd{}{x}+\pd{}{u}, \qquad X_6=t\pd{}{y}+\pd{}{v}, \qquad X_7=t\pd{}{z}+\pd{}{w}, \nonumber\ee
\be X_8=y\pd{}{z}-z\pd{}{y}+v\pd{}{w}-w\pd{}{v}, \qquad X_9=z\pd{}{x}-x\pd{}{z}+w\pd{}{u}-u\pd{}{w}, \nonumber\ee
\be X_{10}=x\pd{}{y}-y\pd{}{x}+u\pd{}{v}-v\pd{}{u}, \nonumber\ee
\be X_{11}=2t\pd{}{t}+x\pd{}{x}+y\pd{}{y}+z\pd{}{z}-u\pd{}{u}-v\pd{}{v}-w\pd{}{w}-2p\pd{}{p}, \nonumber\ee
\be X_{12}=x\pd{}{x}+y\pd{}{y}+z\pd{}{z}+u\pd{}{u}+v\pd{}{v}+w\pd{}{w}-2\rho\pd{}{\rho}. \nonumber\ee
The generic element $T_{ϵ}$ of the one-dimensional Lie symmetry group generated by each of these vector fields can be computed by solving the equations
\begin{equation}
	\begin{cases}
		\td{\hat r}{ϵ}=\xi^r(\hat t,\hat{\x},\hat{\u},\hat p,\hat {ρ}),\quad r=t,x,y,z,\\\\
		\td{\hat q}{ϵ}=\xi_q(\hat t,\hat{\x},\hat{\u},\hat p,\hat {ρ}),\quad q=u,v,w,p,ρ,\\\\
		\hat r(ϵ=0)=r,\\\\
		\hat q(ϵ=0)=q.
	\end{cases}
\end{equation}
These groups combines into the 12-dimensional Lie symmetry group $G$ of equations (\ref{nsc}), generated by the following point transformations:
\begin{itemize}
 \item time translations, obtained from $X_1$:
	 \be (t,\x,\u,p,\rho)\mapsto(t+ϵ,\x,\u,p,\rho)\label{time},\ee
 \item space translations, encoded by $X_2$, $X_3$, and $X_4$:
	 \be (t,\x,\u,p,\rho)\mapsto(t,\x+\vt{ε},\u,p,\rho),\ee
 \item Galilean transformations, corresponding to $X_5$, $X_6$, $X_7$:
       \be (t,\x,\u,p,\rho)\mapsto(t,\x+\vt{ε}t,\u+\vt{ε},p,\rho), \ee
 \item rotations, induced by $X_8$, $X_9$ and $X_{10}$,:
       \be (t,\x,\u,p,\rho)\mapsto(t,\tsr{R}\x,\tsr{R}\u,p,\rho), \ee
 \item and the two scale transformations, generated respectively by $X_{11}$ and $X_{12}$:
	 \be (t,\x,\u,p,\rho)\mapsto(\e^{2ϵ}t,\e^{ϵ}\x,\e^{-ϵ}\u,\e^{-2ϵ}p,\rho), \ee
	 \be (t,\x,\u,p,\rho)\mapsto(t,\e^{ϵ}\x,\e^{ϵ}\u,p,\e^{-2ϵ}\rho). \label{scale2}\ee
\end{itemize}
In these expressions, $ϵ$, $\vt{ε}$ and $\tsr{R}$ are respectively arbitrary scalar, vector and 3D rotation matrix.

Vector fields $X_i$, $i=1,\cdots,12$, constitute a basis of the 12-dimensional Lie algebra $\mathfrak{g}$ of the Lie symmetry group $G$. The commutation table of $\mathfrak{g}$ is presented on Table \ref{commutationtable}. It shows that the subalgebra 
\begin{equation}
	\mathfrak{g}^\text{rad}=\operatorname{span}(X_1,X_2,X_3,X_4,X_5,X_6,X_7,X_{11},X_{12})
\end{equation} is solvable. Indeed, the derived series terminates in the zero algebra:
\be\begin{array}{rll}
	[\mathfrak{g}^\text{rad},\mathfrak{g}^\text{rad}]&=&\operatorname{span}(X_1,X_2,X_3,X_4,X_5,X_6,X_7),\\[5pt]
	[[\mathfrak{g}^\text{rad},\mathfrak{g}^\text{rad}],\mathfrak{g}^\text{rad}]&=&\operatorname{span}(X_2,X_3,X_4),\\[5pt]
	[[[\mathfrak{g}^\text{rad},\mathfrak{g}^\text{rad}],\mathfrak{g}^\text{rad}],\mathfrak{g}^\text{rad}]&=&\{0\}.
\end{array}\ee

\begin{table}\small
$$\begin{array}{c|cccc|ccc|ccc|cc}
 &X_1&X_2&X_3&X_4&X_5&X_6&X_7&X_8&X_9&X_{10}&X_{11}&X_{12}
 \\\hline
X_1&0&0&0&0&X_2&X_3&X_4&0&0&0&2X_1&0
\\
X_2&0&0&0&0&0&0&0&0&\m X_4&X_3&X_2&X_2
\\
X_3&0&0&0&0&0&0&0&X_4&0&\m X_2&X_3&X_3
\\
X_4&0&0&0&0&0&0&0&\m X_3&X_2&0&X_4&X_4
\\\hline
X_5&\m X_2&0&0&0&0&0&0&0&\m X_7&X_6&\m X_5&X_5
\\
X_6&\m X_3&0&0&0&0&0&0&X_7&0&\m X_5&\m X_6&X_6
\\
X_7&\m X_4&0&0&0&0&0&0&\m X_6&X_5&0&\m X_7&X_7
\\\hline
X_8&0&0&\m X_4&X_3&0&\m X_7&X_6&0&\m X_{10}&X_9&0&0
\\
X_9&0&X_4&0&\m X_2&X_7&0&\m X_5&X_{10}&0&\m X_8&0&0
\\
X_{10}&0&\m X_3&X_2&0&\m X_6&X_5&0&\m X_9&X_8&0&0&0
\\\hline
X_{11}&\m2 X_1&\m X_2&\m X_3&\m X_4&X_5&X_6&X_7&0&0&0&0&0
\\
X_{12}&0&\m X_2&\m X_3&\m X_4&\m X_5&\m X_6&\m X_7&0&0&0&0&0
\end{array}$$\caption{Commutation table of $\mathfrak{g}$}\label{commutationtable}
\end{table}

The subalgebra $\mathfrak{g}^\text{s}=\operatorname{span}(X_8,X_9,X_{10})$ of infinitesimal rotations is semi-simple. The Levi decomposition of $\mathfrak{g}$ is then
\be \mathfrak{g}=\mathfrak{g}^\text{rad}\oplus \mathfrak{g}^\text{s}
\ee
$\mathfrak{g}^\text{rad}$ being the radical.

In the next sections, successive reductions are applied to the equations and some self-similar solutions are computed.
We begin with steady  bidimensional solutions.

\section{Steady bidimensional solutions}\label{steady}

In this section, we take $w=0$ and look for solutions invariant under both $X_1$ and $X_4$ that are steady and bidimensional solutions. We write:
\begin{equation*}u(t,x,y)=u_1(x,y), \qquad v(t,x,y)=v_1(x,y), \qquad p(t,x,y)=p_1(x,y),\end{equation*}
\begin{equation} \rho(t,x,y)=\rho_1(x,y). \end{equation}
The reduced equations write:
\begin{eqnarray}\begin{cases}
\pd{\rho_1 u_1}{x}+\pd{\rho_1 v_1}{y}=0 ,
\\[10pt]
\rho_1\left(u_1\pd{u_1}{x}+v_1\pd{u_1}{y}\right) +\pd{p_1}{x}= 
\cfrac{\mu}{3}\left(4\pd{^2u_1}{x^2}+\ppd{v_1}{x}{y} +3\pd{^2u_1}{y^2}\right),
\\[10pt]
\rho_1\left(u_1\pd{v_1}{x}+v_1\pd{v_1}{y}\right) +\pd{p_1}{y}= 
\cfrac{\mu}{3}\left(\ppd{u_1}{x}{y}+4\pd{^2v_1}{y^2} + 3\pd{^2 v_1}{x^2}\right),
\\[10pt]
\cfrac{C_v}{R}\left(\pd{p_1u_1}{x}+\pd{p_1v_1}{y}\right)=\sigma:\tsr{S}+\cfrac{\kappa }{R}
\left(\pd{^2p_1/\rho_1}{x^2}+\pd{^2p_1/\rho_1}{y^2}\right).
\end{cases}\label{eqx4x1}\end{eqnarray}
To find solutions, we reduce these new equations further. The Lie algebra of equations (\ref{eqx4x1}) is spanned by:
\begin{equation}\begin{array}{l}
		Y_1=\pd{}{x}, \hspace{6mm} Y_2=\pd{}{y}, \hspace{6mm} Y_3=x\pd{}{y}-y\pd{}{x}+u_1\pd{}{v_1}-v_1\pd{}{u_1}, \\[10pt]
		Y_4=x\pd{}{x}+y\pd{}{y}-u_1\pd{}{u_1}-v_1\pd{}{v_1}-2p_1\pd{}{p_1}, \\[10pt]
Y_5=x\pd{}{x}+y\pd{}{y}+u_1\pd{}{u_1}+v_1\pd{}{v_1}-2\rho_1\pd{}{\rho_1}. 
\end{array}\end{equation}

\subsection{Reduction with $Y_1$}

Reduction under vector field $Y_1$ suggests a solution in the form
\begin{eqnarray}u_1(x,y)=u_2(y), \qquad v_1(x,y)=v_2(y), \\\\ p_1(x,y)=p_2(y), \qquad \rho_1(x,y)=\rho_2(y).\end{eqnarray}
Inserting these expressions in equations (\ref{eqx4x1}), it follows:
\begin{equation}\begin{cases}
 (\rho_2v_2)'=0, \\
 \rho_2v_2u_2'=\mu u_2'',\\
 3\rho_2v_2v'_2+p_2'=4\mu v_2',\\
 3C_v(p_2v_2)'=R(4\mu v_2'^2-3v_2'p_2+3\mu u_2'^2)+3\kappa (p_2/\rho_2)''.
\end{cases}\label{x2x1y1}\end{equation}

In equations (\ref{x2x1y1}) and in the rest of the document, a prime symbol is used to designate a derivation of a function depending on a single variable. One solution of (\ref{x2x1y1}) is
\begin{equation}\rho_2(y)=\cfrac{a}{v_2(y)},\qquad u_2(y)=u_3+u_4\e^{\frac{ay}{\mu}},\qquad v_2(y)=v_3\e^{\frac{ay}{\mu}}, \qquad p_2(y)=p_3\e^{\frac{ay}{\mu}}\end{equation}
where $a$, $u_3$, $u_4$ and $p_3$ are constants linked by the relations
\begin{equation}p_3=\cfrac{av_3}{3} \eqspc{and} u_4^2=\left(\cfrac{2C_v}{3R}-\cfrac{4\kappa}{3R\mu}-1\right)v_3^2.\end{equation}
We deduce the following class of solutions of (\ref{nsc}):
\begin{equation}\begin{mybox}\\[-10pt]\displaystyle 
u(t,x,y)=u_3\pm\sqrt{\cfrac{2C_v\mu-4\kappa-3R\mu}{3R\mu}}\ v_3\e^{\frac{ay}{\mu}}, \qquad v(t,x,y)=v_3\e^{\frac{ay}{\mu}}, 
\\[10pt]\displaystyle 
p(t,x,y)=\cfrac{av_3}{3}\e^{\frac{ay}{\mu}}, \qquad \rho(t,x,y)=\cfrac{a}{v_3}\e^{-\frac{ay}{\mu}}. \\[-10pt]
\end{mybox}\label{solx1x2x4}\end{equation}
It can be observed that the pressure and the density are respectively proportional and inversly proportional to $v$.

\begin{figure}[ht]\centering
	\includegraphics[width=37mm]{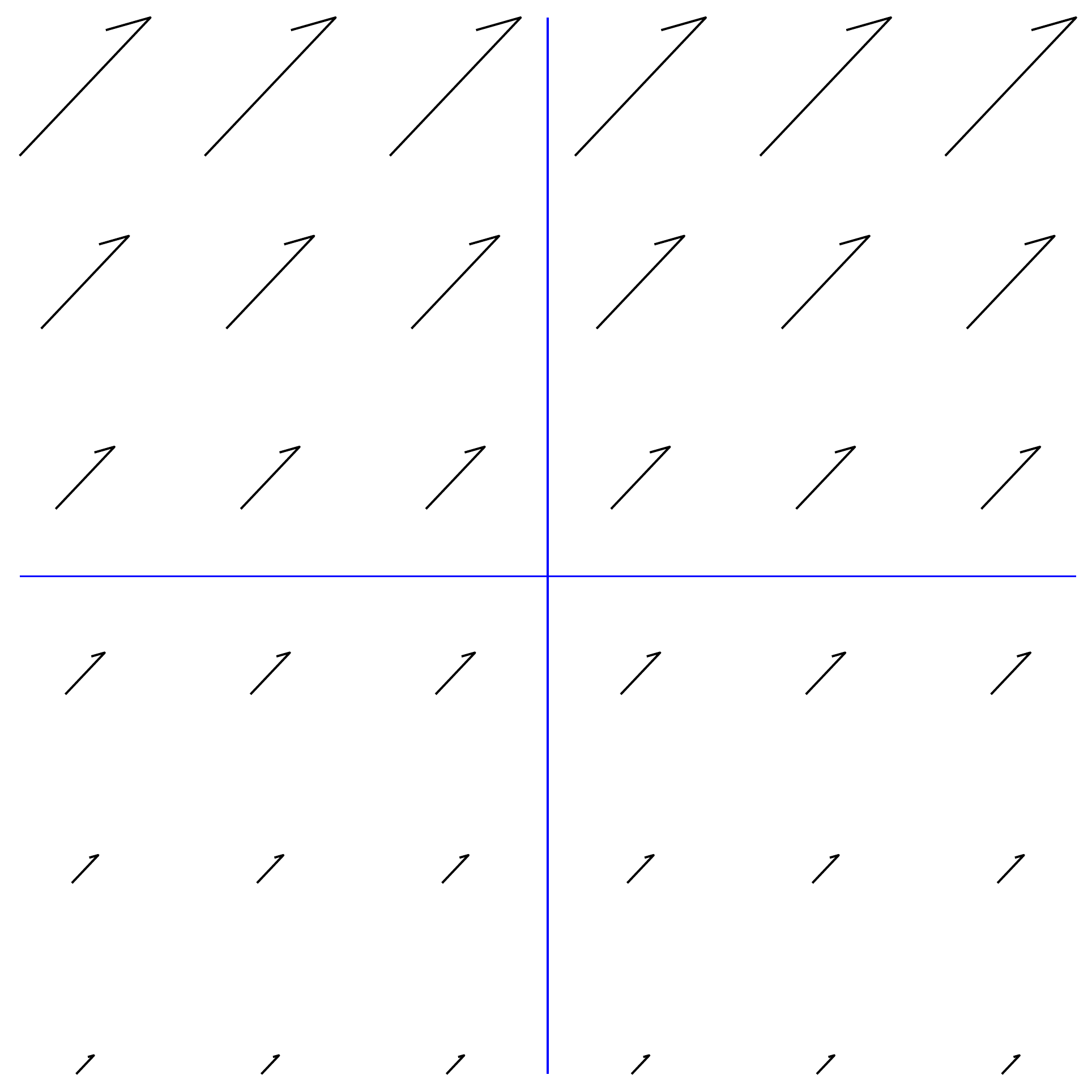}\hfill
	\includegraphics[width=37mm]{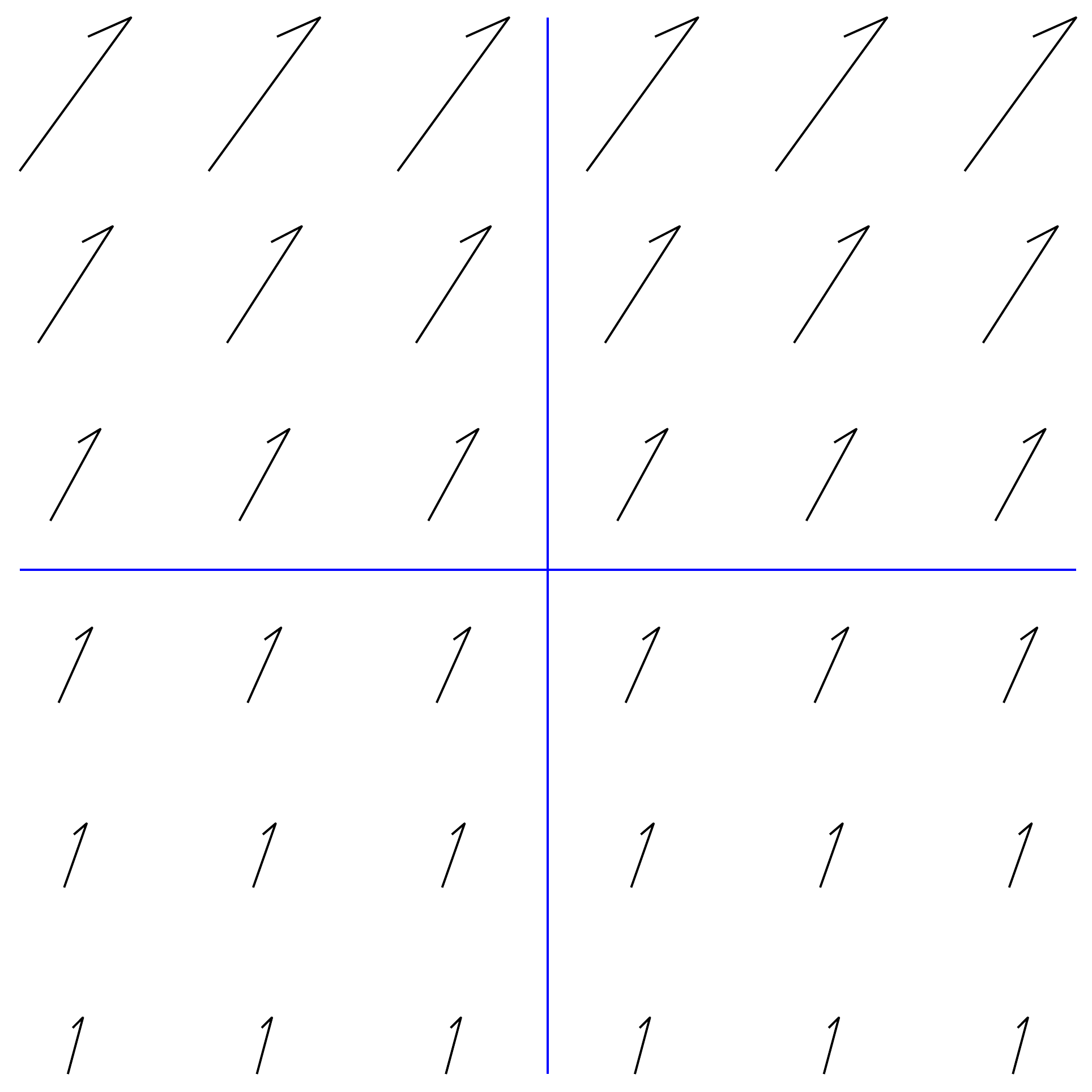}\hfill
	\includegraphics[width=37mm]{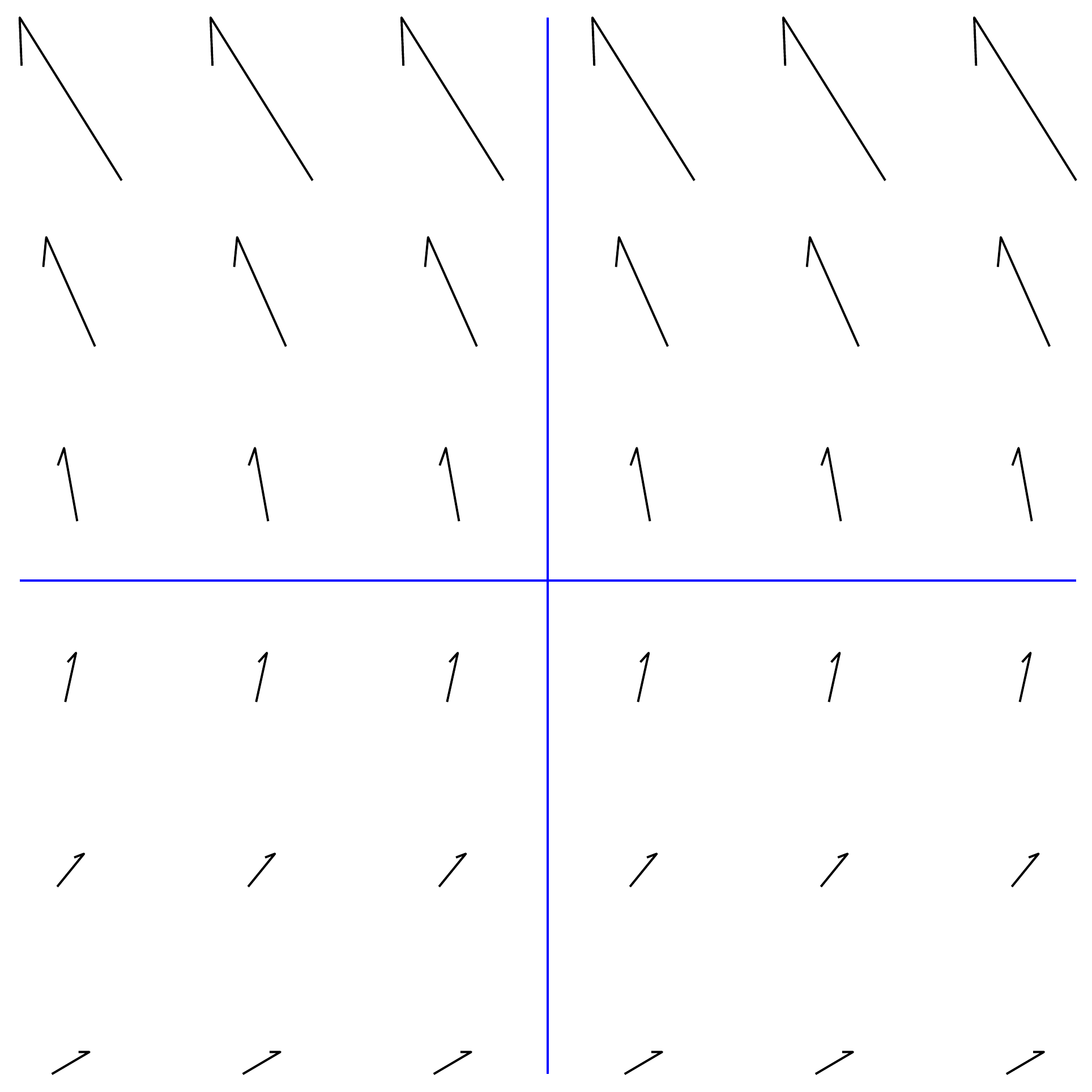}
	\caption{Left: $\u=(\e^y,\e^y)$. Center:  $\u=(1+\e^y,\e^y)$. Right: $\u=(1-\e^y,\e^y)$}
	\label{figx1x2x4}
\end{figure}

When $u_3=0$, the flow is parallel, as can be observed in Figure \ref{figx1x2x4}, left. Indeed, in a suitable orthogonal frame,
$$u=u_4\e^{\frac a\mu(cx+dy)},\hspc v=0$$
for some constants $u_4$, $c$ and $d$. This solution exhibits an exponential growth of norm of the velocity in both $x$ and $y$ directions. 

Some other simple solutions belonging to class (\ref{solx1x2x4}) are plotted in Figure \ref{figx1x2x4}.

\subsection{Reduction with $Y_3$}

\begin{figure}\centering
\includegraphics[width=4cm]{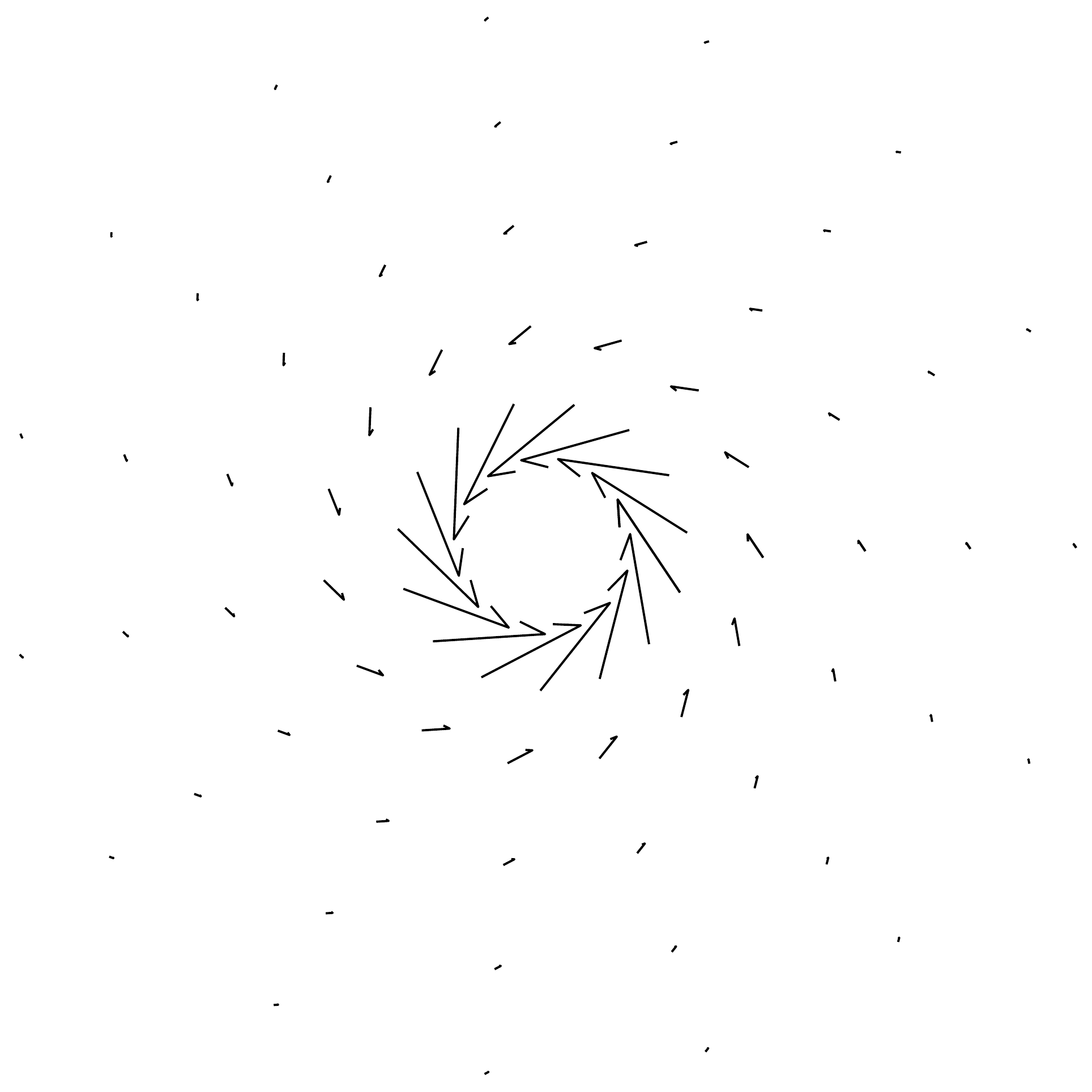} 
\caption{Steady vortex sink. $‖\u‖\propto r^{-2}$}\label{vortex_sink}
\end{figure}

Self-similar solutions under the infenitesimal rotation $Y_3$ write:
\begin{equation} u_1=u_2(r)\cos \theta-v_2(r)\sin \theta, \qquad v_1=u_2(r)\sin \theta+v_2(r)\cos \theta, 
\nonumber\end{equation} \begin{equation}p_1=p_2(r),\qquad \rho_1=\rho_2(r) \end{equation}
where $(r,\theta)$ are the polar coordinates. The equations are reduced into
\begin{equation}\begin{cases}
 (r\rho_2 u_2)'=0
 \\[5pt]
\rho_2\left(u_2u'_2-\cfrac{v_2^2}{r}\right)+p_2'=\cfrac{4\mu}{3} \left(u_2''+\cfrac{u_2'}{r}-\cfrac{u_2}{r^2}\right)
\\[15pt]
\rho_2\left(u_2v_2'-\cfrac{u_2v_2}{r}\right)=\mu
\left(v_2''+\cfrac{v_2'}{r}-\cfrac{v_2}{r^2}\right)
\\[15pt]
\cfrac{Cv}{R}\left((p_2u_2)'+\cfrac{p_2u_2}{r}\right)=E+\cfrac{\kappa}{R} \left( (p_2/\rho_2)''+\cfrac{(p_2/\rho_2)'}{r}\right)
\end{cases}\label{x2x1y2}\end{equation}
with
\begin{equation}E= -\cfrac{p_2(ru_2)'}{r}+\cfrac{\mu}{3r^2} \left( 4u_2^2+3v_2^2-4ru_2u_2'-6rv_2v_2'+4r^2u_2'^2+3r^2v_2'^2 \right). \nonumber\end{equation}
These equations admit the following infinitesimal symmetry
\begin{equation}
 r\pd{}{r}-2u_2\pd{}{u_2}-2v_2\pd{}{v_2}+\rho_2\pd{}{\rho_2}-3p\pd{}{p_2}.
\end{equation}
A self-similar solution under this symmetry verify:
\begin{equation}
u_3=r^2u_2(r), \qquad v_3=r^2v_2(r), \qquad p_3=r^3p_2(r), \qquad \rho_3=\cfrac{\rho_2(r)}{r}.
\label{varx2x1y2}\end{equation}
where $u_3,v_3,p_3$ and $ρ_3$ are constants. Inserting (\ref{varx2x1y2}) into equations (\ref{x2x1y2}) leads to the following algebraic relations on these constants:
\begin{equation}
 \begin{cases} 
\rho_3u_3=-\mu\\
\rho_3v_3^2+3p_3+2\mu u_3=0\\
3p_3(4C_v\mu+R\mu-16\kappa)=\mu R\rho_3(28u_3^2+27v_3^2).
  \label{eqx2x1y22}\end{cases}
\end{equation}
The velocity components verify:
\[u(r,\theta)=\cfrac{u_3\cos \theta-v_3\sin \theta}{r^2}, \quad
v(r,\theta)=\cfrac{u_3\sin \theta+v_3\cos \theta}{r^2}.\]
If $\ve_r$ and $\ve_\theta$ designate the unitary radial and angular vectors, then the solution, in polar coordinates, is
\begin{equation}
	\begin{mybox} \vt u(r,\theta)=\cfrac{1}{r^2}(u_3\ve_r+v_3\ve_\theta),\hspc p(r,\theta)=\cfrac{p_3}{r^3}, \hspc \rho(r,\theta)=\rho_3r \\[-10pt]\end{mybox}
\label{solx2x1y22}\end{equation}
This solution represents a steady vortex flow. It is sketched in Figure \ref{vortex_sink}. The constant $u_3$ beeing negative, the origin is a sink. The sign of $v3$ determines the direction of rotation. The velocity magnitude increases as $r^{-2}$ towards the sink. 

\subsection{Reduction with $Y_4$}

\begin{figure}[ht]\centering
\includegraphics[width=4cm]{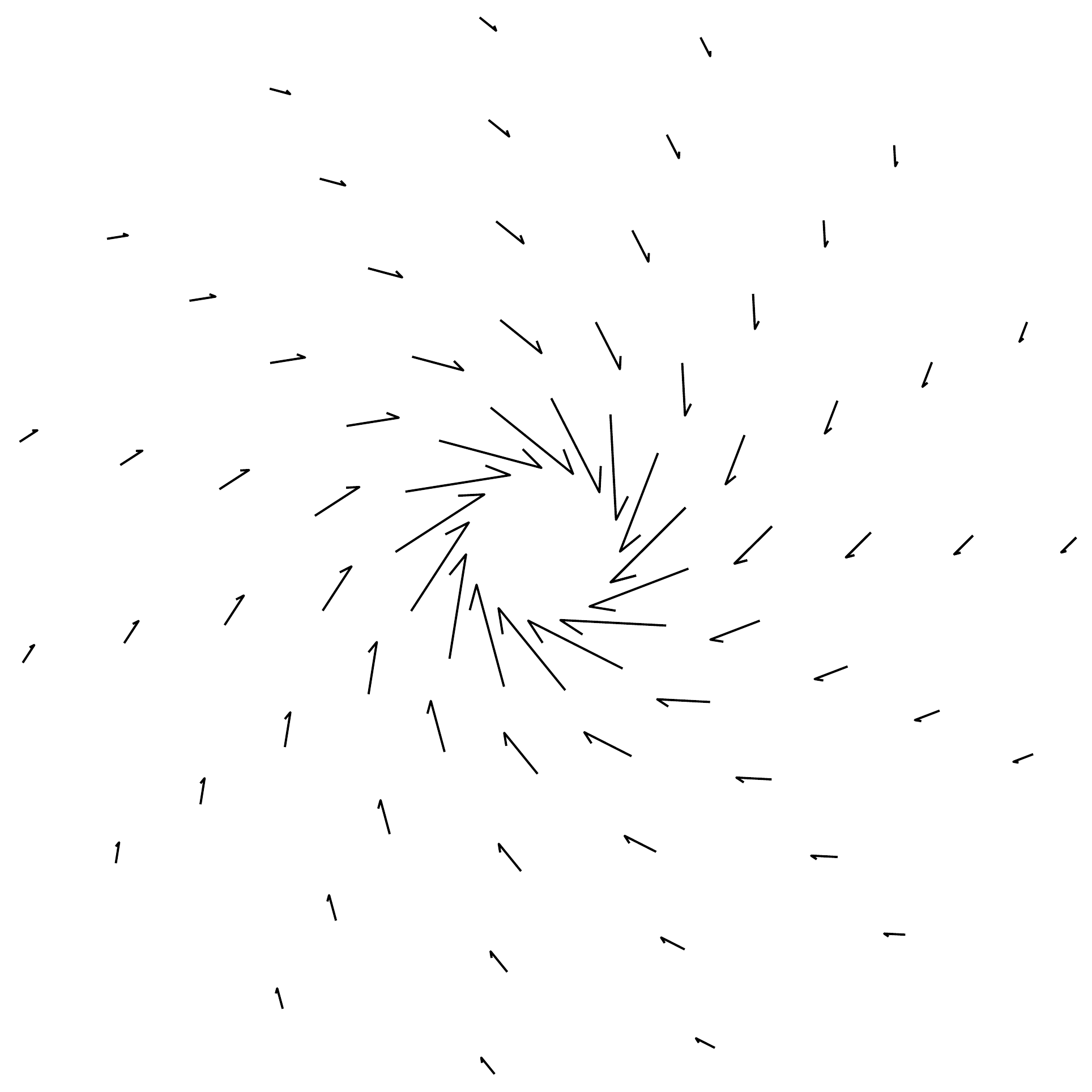}\qquad\quad
\includegraphics[width=4cm]{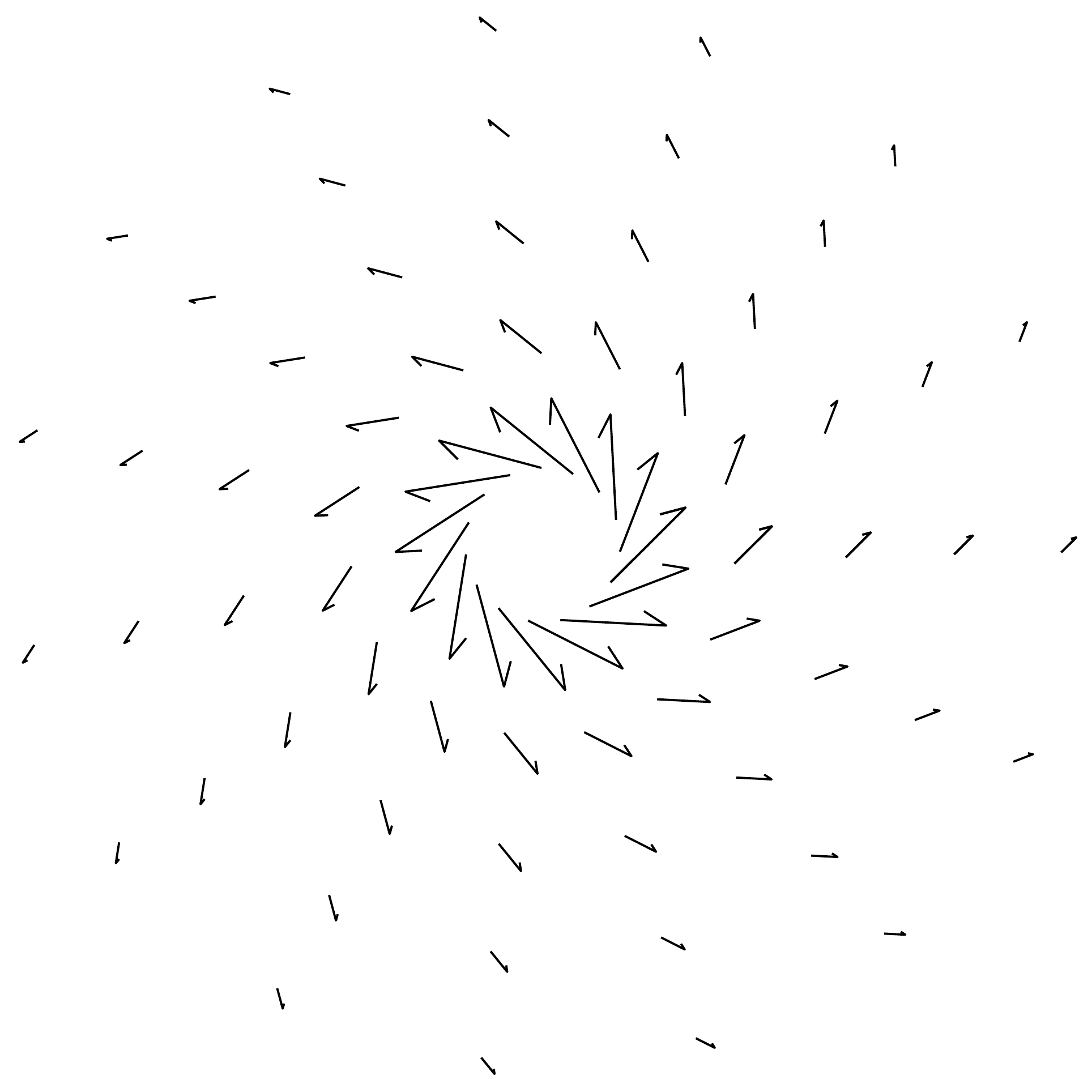} 
\caption{Vortex sink (left) and source (right). $‖\u‖\propto r^{-1}$}\label{vortex_r1}
\end{figure}

A basis of invariants under the scale transformation generated by $Y_4$ is
\begin{equation} \eta=\cfrac{y}{x}, \hspc u_2(\eta)=xu_1(x,y), \hspc v_2(\eta)=xv_1(x,y),\nonumber \end{equation}
\begin{equation} p_2(\eta)=x^2p_1(x,y), \hspc \rho_2(\eta)=\rho_1(x,y).\end{equation}
The reduced equations are
\begin{equation}
	\begin{cases}
	( ρ_2v_2-ρ_2u_2η)'=0
	\\ \displaystyle 
	ρ_2⦅v_2u'_2-u_2(u_2η)'⦆-p_2'η-2p_2+÷{µ}3⦅v_2η-3u_2-4u_2η^2⦆''=0
	\\ \displaystyle 
	ρ_2⦅v_2v'_2-u_2(v_2η)'⦆+p_2'+÷{µ}3⦅u_2η-4v_2-3v_2η^2⦆''=0
	\\ \displaystyle 
	C_v⦅p_2(v_2-u_2η)⦆'-2C_vp_2u_2+Rp_2(v_2-u_2η)'-Rµ\tsr S_2+
	\\\displaystyle
	\qquad\qquad\qquad\qquad\qquad κ⦅⦅÷{p_2}{ρ_2}⦆''η^2+6⦅÷{p_2}{ρ_2}⦆'η+6÷{p_2}{ρ_2}⦆=0
	\end{cases}
	\label{eqx1x4y4}
\end{equation}
where
\[ \tsr S_2=÷43(v'_2-u'_2η-u_2)^2+÷43v'_2u_2+(u'_2-v'_2η-v_2)^2.\]
The solution of system (\ref{eqx1x4y4}) is
\begin{equation}
 u_2(\eta)=\cfrac{(\rho_3-v_3\eta)u_3}{\rho_3(1+\eta^2)},\hspc v_2(\eta)=u_2(\eta)\eta+\cfrac{v_3}{\rho_2(\eta)},  \nonumber\end{equation}
\begin{equation} p_2(\eta)=\cfrac{(\rho_3^2+v_3^2)u_2(\eta)}{2(v_3\eta-\rho_3)}, \hspc \rho_2(\eta)=\cfrac{\rho_3-v_3\eta}{(1+\eta^2)u_2(\eta)}
\end{equation}
where $u_3$ and $v_3$ are constants and 
\begin{equation}\rho_3=\cfrac{-2\kappa+4\mu R}{C_v}.\end{equation}
As a result,
\begin{equation}\begin{mybox}\\
u(t,x,y)=\cfrac{(\rho_3x-v_3y)u_3}{(x^2+y^2)\rho_3}, \hspc v(t,x,y)=\cfrac{(\rho_3y+v_3x)u_3}{(x^2+y^2)\rho_3}, \\\\ p(t,x,y)=\cfrac{-(\rho_3^2+v_3^2)u_3}{2(x^2+y^2)\rho_3}, \hspc \rho(t,x,y)=\cfrac{\rho_3}{u_3}
\\\end{mybox}\end{equation}
This is an incompressible solution, representing also a vortex, but with a velocity magnitude proportional to $r^{-1}$ towards the origin. If $u_3>0$, it models a swirling source and when $u_3<0$, the origin is a sink. These two cases are represented in Figure \ref{vortex_r1} when $|u_3|=1$ and $v_3/ρ_3=1$.

\subsection{Reduction with a linear combination of $Y_1$ and $Y_2$}

\begin{figure}\centering
\includegraphics[width=4cm]{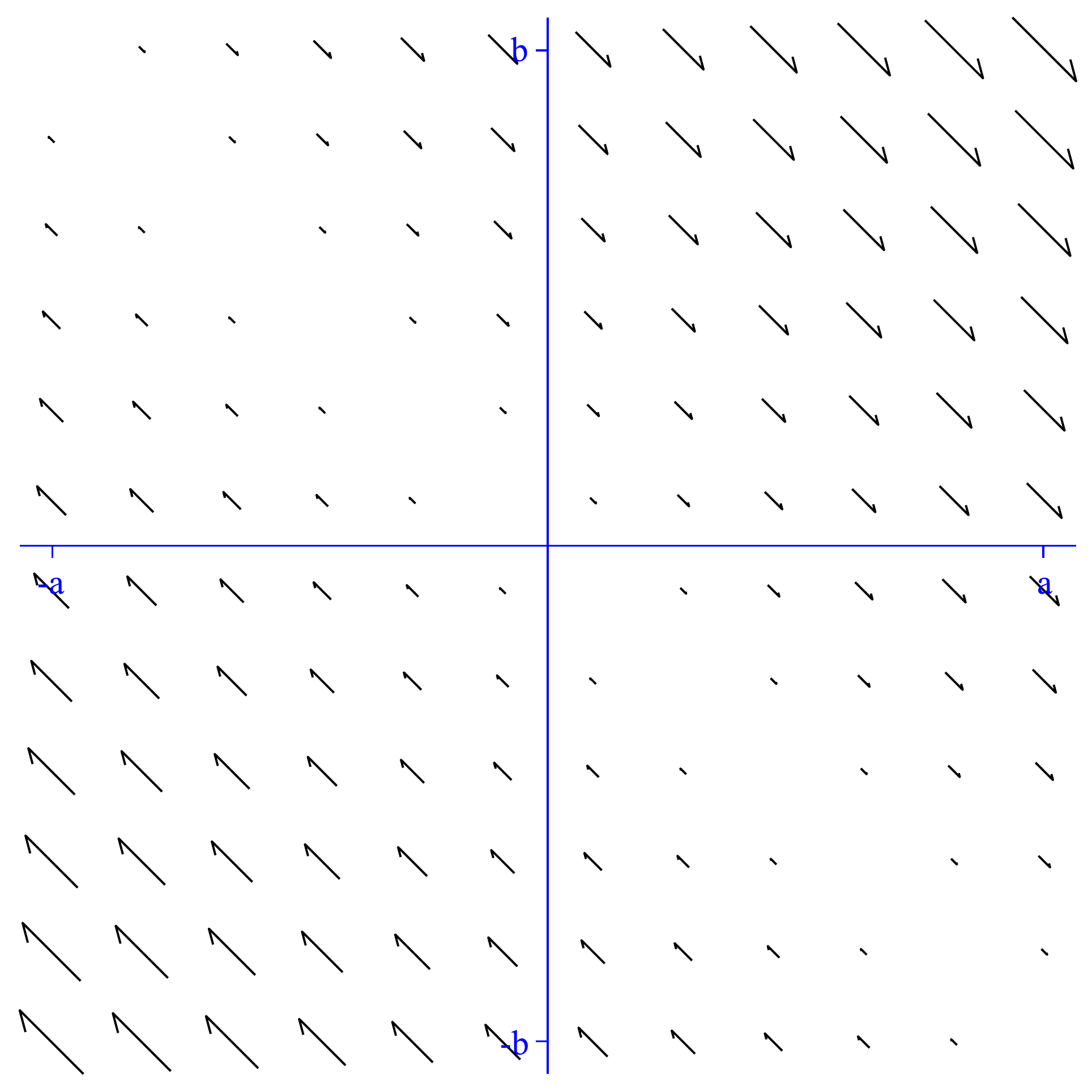} 
\caption{Parallel flow}\label{parallel}
\end{figure}

Invariant solutions under a linear combination $-bY_1+aY_2$, for some constants $a$ and $b$, can be written as follows:
\begin{eqnarray}
	u_1(x,y)=u_2(\eta), \qquad v_1(x,y)=v_2(\eta), \\[5pt]p_1(x,y)=p_2(\eta), \qquad  \rho_1(x,y)=\rho_2(\eta),
\label{mby1ay2}\end{eqnarray}
where the self-similarity variable is 
\[
η=ax+by.
\]
Relations (\ref{mby1ay2}) transform equations (\ref{x2x1y2}) into
\begin{equation}
 \begin{cases}
a(\rho_1u_1)'+b(\rho_1v_1)'=0
\vspace{5pt}\\ 
3\rho_1\left( au_1u_1'+bv_1u_1'\right) +3ap_1'= \mu \big( (4a^2+3b^2)u_1''+abv_1''\big)
\vspace{5pt}\\ 
3\rho_1\left(au_1v_1'+bv_1v_1'\right) +3bp_1'= \mu \big((4b^2+3a^2)v_1''+abu_1''\big)
\vspace{4pt}\\ 
\cfrac{C_v}{R}\big(a(p_1u_1)'+b(p_1v_1)'\big) =\sigma:\tsr{S} +(a^2+b^2)\cfrac{\kappa}{R} (p_1/\rho_1)''
\end{cases}
\end{equation}
One solution can easily be found if $u_2$ and $v_2$ are linear. In this case
\begin{eqnarray}
 u_2(\eta)=u_3\eta, \qquad v_2(\eta)=\cfrac{-au_3\eta}{b}, \qquad p_2(\eta)=p_3\\\\
\rho_2(\eta)=\cfrac{-2\kappa p_3b^2}{\mu u_3^2R\eta^2(a^2+b^2)+2\kappa p_3(\rho_3-\rho_4\eta^2)}.
\end{eqnarray}
where $u_3$, $p_3$, $\rho_3$ and $\rho_4$ are constants. We get.
\begin{equation}
 \begin{mybox}
	 u(t,x,y)=(ax+by)u_3, \quad v(t,x,y)=-\cfrac{a}{b}(ax+by)u_3, \quad p(t,x,y)=p_3, \\[10pt]
\rho(t,x,y)=\cfrac{-2\kappa p_3b^2}{\mu u_3^2R(ax+by)^2(a^2+b^2)+2\kappa p_3(\rho_3-\rho_4(ax+by)^2)}, \\[-10pt]\end{mybox}
\end{equation}
This solution represents a parallel, but direction-changing flow, with a uniform pressure. It is represented in Figure \ref{parallel} in the case $a>0$ and $b>0$. When $a=0$ the flow is parallel to $x$-axis.

In the next section, we seek usteady solutions.

\section{Unsteady bidimensional solutions}\label{unsteady}

In the case of unsteady bidimensional flow, the equations are
\begin{eqnarray}\begin{cases}
\pd{\rho}{t}+\pd{\rho u}{x}+\pd{\rho v}{y}=0 
\\[10pt]
\rho\left(\pd{u}{t}+u\pd{u}{x}+v\pd{u}{y}\right) +\pd{p}{x}= 
\cfrac{\mu}{3}\left(4\pd{^2u}{x^2}+\ppd{v}{x}{y} +3\pd{^2u}{y^2}\right)
\\[10pt]
\rho\left(\pd{v}{t}+u\pd{v}{x}+v\pd{v}{y}\right) +\pd{p}{y}= 
\cfrac{\mu}{3}\left(\ppd{u}{x}{y}+4\pd{^2v}{y^2} + 3\pd{^2 v}{x^2}\right)
\\[10pt]
\cfrac{C_v}{R}\left(\pd{p}{t}+\pd{pu}{x}+\pd{pv}{y}\right)=\sigma:\tsr{S}+\cfrac{\kappa }{R}
\left(\pd{^2p/\rho}{x^2}+\pd{^2p/\rho}{y^2}\right)
\end{cases}
\label{unsteady_bidim}
\end{eqnarray}

The Lie algebra of these equations is spanned by $X_1$, $X_2$, $X_3$, $X_5$, $X_6$, $X_{10}$,  $X_{11}$
and $X_{12}$ (without the terms in $\pd {}z$ and $\pd {}w$). Let us begin with solutions homogeneous in $x$ direction, {\it i.e.} invariant under $X_2$.

\subsection{Reduction with $X_2$}

$X_2$-invariant solutions are of the form
\begin{equation}u(t,x,y)=u_1(t,y), \qquad v(t,x,y)=v_1(t,y), \nonumber\end{equation} \begin{equation}p(t,x,y)=p_1(t,y), \qquad \rho(t,x,y)=\rho_1(t,y).\end{equation}
In this case, equations (\ref{unsteady_bidim}) reduce into
\begin{eqnarray}\begin{cases}
\pd{\rho_1}{t}+\pd{\rho_1 v_1}{y}=0 
\\[10pt]
\rho_1\left(\pd{u_1}{t}+v_1\pd{u_1}{y}\right)\! = 
\mu\pd{^2u_1}{y^2}
\\[10pt]
\rho_1\left(\pd{v_1}{t}+v_1\pd{v_1}{y}\right)\! =-\pd{p_1}{y}+
\cfrac{4\mu}{3}\ \pd{^2v_1}{y^2}
\\[10pt]
\cfrac{C_v}{R}\!\left(\!\pd{p_1}{t}+\pd{p_1v_1}{y}\!\right)\!\!=\!-p_1\pd{v_1}{y} 
\!+\!\cfrac{4\mu}{3}\!\left(\!\pd{v_1}{y}\!\!\right)^2 \!\!+\!\mu\!\left(\!\pd{u_1}{y}\!\!\right)^2\!\! +\!\cfrac{\kappa }{R}
\pd{^2(p_1/ρ_1)}{y^2}
\end{cases}\label{eqx2}\end{eqnarray}

These equations admit the following infinitesimal symmetries
\[Z_1=\pd{}{t}, \hspc Z_2=\pd{}{y},  \hspc Z_3=\pd{}{u_1},   \hspc Z_4=t\pd{}{y}+\pd{}{v_1}\]
\[Z_5=2t\pd{}{t}+y\pd{}{y}-u_1\pd{}{u_1}-v_1\pd{}{v_1}-2p_1\pd{}{p_1}, \]
\[Z_6=y\pd{}{y}+u_1\pd{}{u_1}+v_1\pd{}{v_1}-2\rho_1\pd{}{\rho_1}.\]

\begin{figure}[ht]\centering
	\includegraphics[width=4cm]{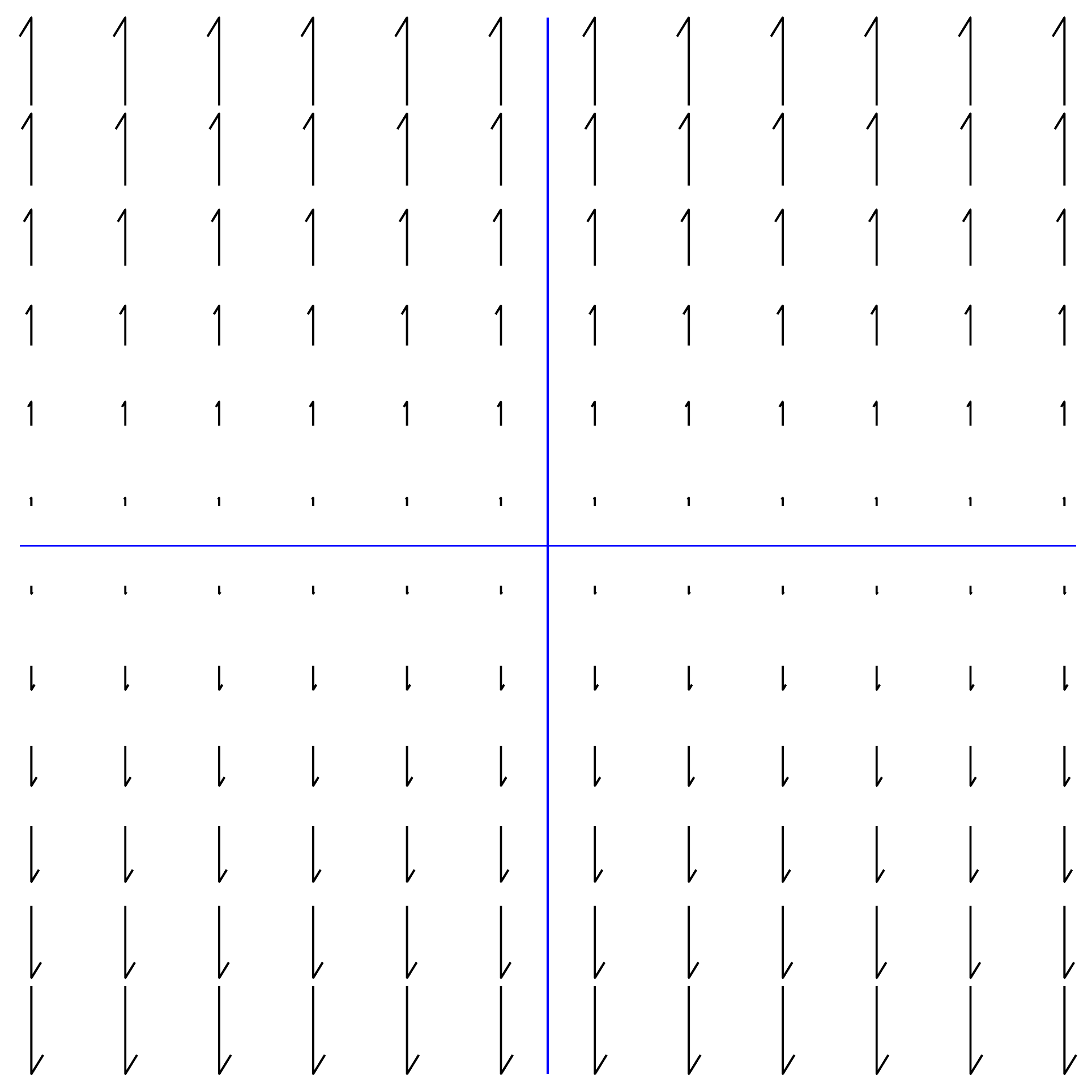}\qquad\quad
	\includegraphics[width=4cm]{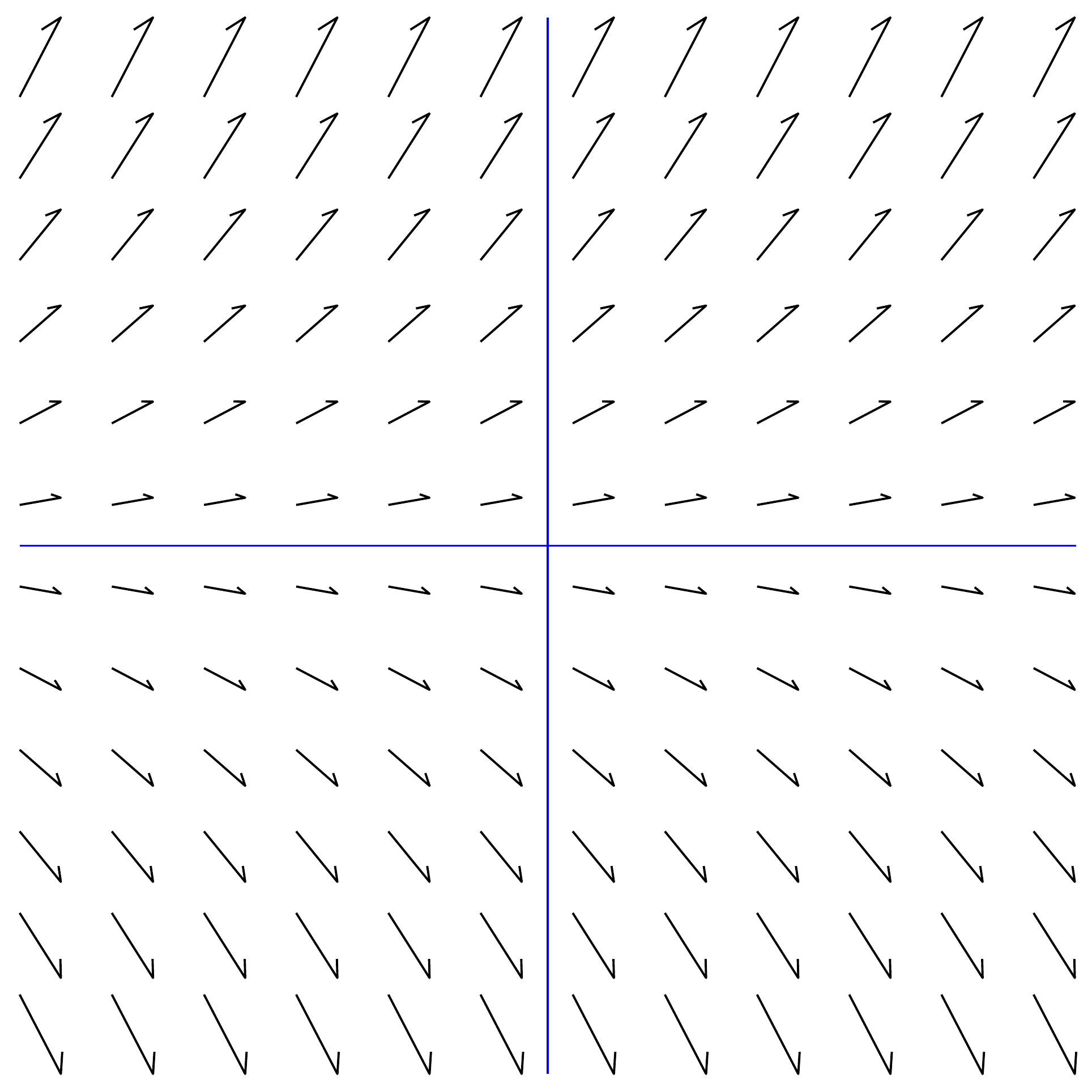}
	\caption{Solution (\ref{eq:unsteady_2d_y5}). Left: $u_3=0$. Right $u_3>0$}\label{unsteady_2d_y5}
\end{figure}

A basis of invariants under $Z_4$ is
 \begin{equation}v_2(t)=v_1(t,y)-\cfrac{y}{t}, \qquad u_1(t,y)=u_2(t), \nonumber\end{equation}\begin{equation} p_1(t,y)=p_2(t), \qquad \rho_1(t,y)=\rho_2(t).\end{equation}
The corresponding solution for $t>0$ is
\begin{equation}\begin{mybox}
		u(t,x,y)=u_3, \qquad\qquad v(t,x,y)=\cfrac{v_3+y}{t}, \\[5pt] p(t,x,y)=\cfrac{4\mu+p_3\ t^{-R/C_v}}{3t}, \qquad
\rho(t,x,y)=\cfrac{\rho_3}{t}, \\[-10pt]\end{mybox}\label{eq:unsteady_2d_y5}\end{equation}
where $u_3$, $v_3$, $p_3$ and $\rho_3$ are constants. The pressure and density are uniform but time-dependent. The flow is sketched in Figure \ref{unsteady_2d_y5} for a fixed $t>0$ and $v_3=0$. $x$-axis can be seen as a wall. When $u_3=0$, the flow is parallel to $y$-axis. 

\begin{figure}[ht]\centering
	\includegraphics[width=4cm]{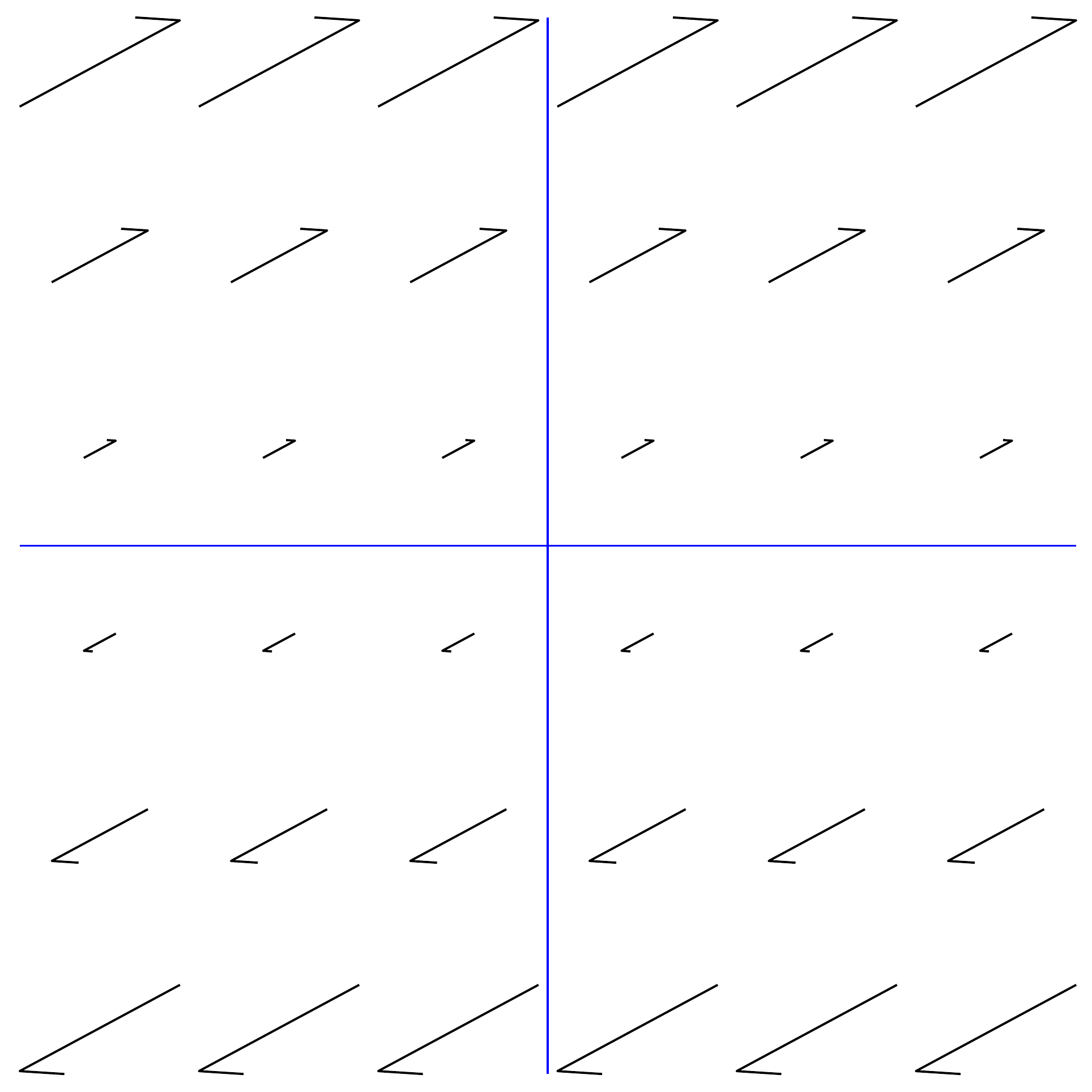}\qquad\quad
	\includegraphics[width=4cm]{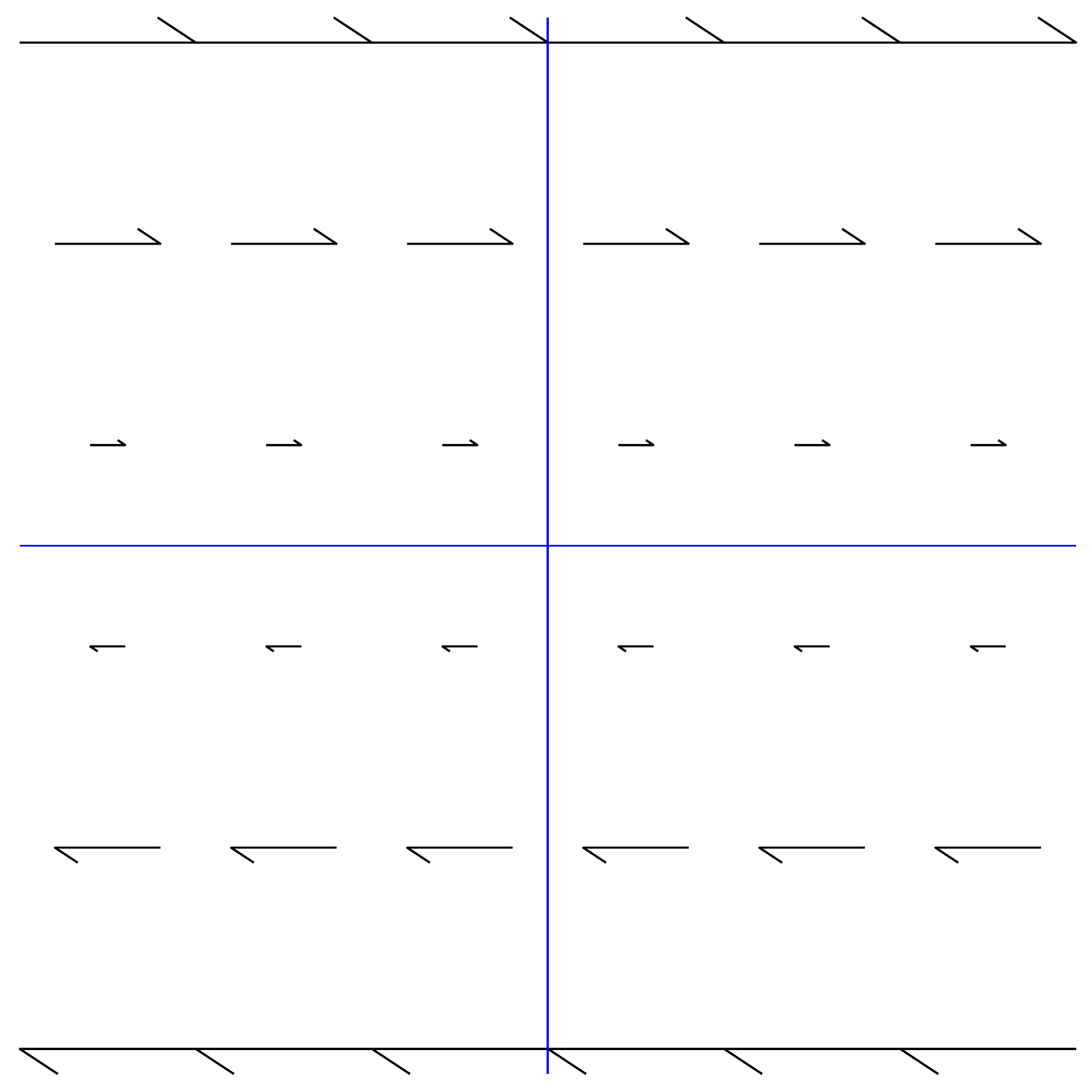}

	\caption{Left: solution (\ref{eq:unsteady_2d_x2y7}). Right: solution (\ref{eq:unsteady_2d_x2y8})}\label{unsteady_2d_x2y7-y8}
\end{figure}

Invariants under the vector field $Z_5$ are
\begin{equation}
 \eta=\cfrac{y}{\sqrt{t}}, \qquad u_2(\eta)=yu_1(t,y), \qquad v_2(\eta)=yv_1(t,y), \nonumber\end{equation}\begin{equation} p_2(\eta)=y^2p_1(t,y), \qquad\rho_2(\eta)=\rho_1(t,y). 
\end{equation}
These invariants reduce equations (\ref{eqx2}) into
\[
	\begin{cases}
		-\eta^3\rho_2'+2(v_2\rho_2'+v_2'\rho_2)\eta-2\rho_2v_2=0 \\[5pt]
		-\eta^3\rho_2u'_2+2\rho_2v_2(u_2'\eta-u_2)+2\mu(-\eta^2u''_2+2\eta u'_2-2u_2)=0 \\[5pt]
		-\eta^3\rho_2v_2'+2\rho_2(v_2v'_2\eta-v_2^2)+2\eta p_2'-4p_2+\cfrac{8\mu}{3}(-\eta^2v_2''+2\eta v_2'-2v_2)=0 \\[5pt]
 \cfrac{C_v}{R}\big( -\eta^3p_2'+2v_2p'_2\eta-6v_2p_2 +2\eta p_2v_2'\big)   =  2p_2\big(-\eta v'_2+ v_2\big)
\\
\qquad  +2\mu\big(\eta^2u_2'^2-2\eta u_2u'_2+u_2^2 \big) +\cfrac{8\mu}{3} \big(  \eta^2v_2'^2 -2\eta v_2'v_2+v_2^2\big) 
+\cfrac{2\kappa\eta^4}{R}\bigg( \cfrac{p_2}{\eta^2\rho_2}\bigg)''
\end{cases}
\]
A solution of these equations can be found with the ansatz:
\begin{equation}
 u_2(\eta)=u_3\eta^2, \qquad v_2(\eta)=v_3\eta^2, \qquad p_2(\eta)=p_3\eta^2, \qquad \rho_2=\rho_3\eta^{-2}
\end{equation}
where $u_3$, $v_3$, $p_3$ and $ρ_3$ are constants.
Inserting these relations into the equations implies relations on theses constants and leads to the following parallel flow:
\begin{equation}\begin{mybox}
		u(t,x,y)=\cfrac{u_3y}{t}, \qquad\qquad v(t,x,y)=\cfrac{y}{t}, \\[10pt] p(t,x,y)=\cfrac{\mu\rho_3R(4+3u_3^2)}{3t(\rho_3R-2\kappa)}, \qquad
\rho(t,x,y)=\cfrac{\rho_3t}{y^2}. 
\\[-10pt]\end{mybox}\label{eq:unsteady_2d_x2y7}\end{equation}
The flow is plotted in Figure \ref{unsteady_2d_x2y7-y8}, left.

Self-similar solutions of equations (\ref{eqx2}) under $Z_6$ have the form
\begin{equation} u_1(t,y)=yu_2(t), \qquad v_1(t,y)=yv_2(t), \nonumber\end{equation}\begin{equation}
p_1(t,y)=p_2(t),\qquad \rho_1(t,y)=\rho_2(t)/y^2.\end{equation}
This change of variables simplifies the equations into
\begin{equation}\begin{cases}
-\rho_2'+v_2\rho_2=0\\
u_2'+u_2v_2=0\\
v_2'+v_2^2=0\\
\cfrac{3C_v}{R} (p_2'+p_2v_2)=-3p_2v_2+\mu(3u_2^2+4v_2^2)+\cfrac{6\kappa p_2}{R\rho_2}
\end{cases}\label{eqx2y8}\end{equation}
One solution of equations (\ref{eqx2y8}) is
\begin{equation}
 u_2(t)=u_3, \qquad v_2(t)=0, \qquad \rho_2(t)=\rho_3, \nonumber\end{equation}\begin{equation} p_2(t)=p_3\exp\left(\frac{2\kappa t}{\rho_3C_v}\right)  -\cfrac{\rho_3\mu Ru_3^2}{2\kappa},
\end{equation}
where $u_3$, $p_3$ and $\rho_3$ are arbitrary constants. We get the following self-similar solution under $X_2$ and $Z_6$:
\begin{equation}
	\begin{mybox}\\[-10pt]
	 u(t,x,y)=u_3y, \hspc p(t,x,y)=p_3\exp\left(\cfrac{2\kappa t}{\rho_3C_v}\right) -\cfrac{\rho_3\mu Ru_3^2}{2\kappa}, \\[10pt]
v(t,x,y)=0, \hspc \rho(t,x,y)=\cfrac{\rho_3}{y^2}. 
\\[-10pt]\end{mybox}\label{eq:unsteady_2d_x2y8}
\end{equation}
This solution is graphically presented in Figure \ref{unsteady_2d_x2y7-y8}, right. If one limits to $y∈[0,h]$ for some constant $h$, it may represent a Couette flow, with a uniform but time-dependent pressure field and a density depending on the wall distance.

Another solution of (\ref{eqx2y8}) is
\[
u_2(t)=\cfrac{u_3}{t+v_3},\qquad v_2=\cfrac{1}{t+v_3},\qquad \rho_2(t)=(t+v_3)\rho_3, \]
\[ p_2(t)=p_3\exp\left(\cfrac{2\kappa-\rho_3(C_v+R)}{\rho_3C_v}\ln (t+v3)\right) +\cfrac{\rho_3R\mu(4+3u_3^2)}{3(t+v_3)(\rho_3R-2\kappa)}.
\]
where $u_3$, $p_3$ and $\rho_3$ are constants. Hence,
\begin{equation}
 \begin{mybox}
u(t,x,y)=\cfrac{u_3y}{t+v_3}, \qquad
v(t,x,y)=\cfrac{y}{t+v_3}, \qquad\rho(t,x,y)=\cfrac{(t+v_3)\rho_3}{y^2}, \\[10pt]
p(t,x,y)=p_3\exp\!\left(\!\cfrac{2\kappa-\rho_3(C_v+R)}{\rho_3C_v}\ln (t+v_3)\!\!\right) \!+\!\cfrac{\rho_3R\mu(4+3u_3^2)}{3(t+v_3)(\rho_3R-2\kappa)}.
\\[-10pt]\end{mybox}\end{equation}
This solution has the same profile as solution (\ref{eq:unsteady_2d_x2y7}) which is plotted in the left part of Figure \ref{unsteady_2d_x2y7-y8}, but with an algebraic time evolution of the pressure instead of a hyperbolic one.

Still in the bidimensional case, we reduce equations (\ref{unsteady_bidim}) under infinitesimal Galilean transformation $X_5$ and under infinitesimal scale transformations.

\subsection{Galilean transformation}

Self-similar solutions of (\ref{unsteady_bidim}) under $X_5$ can be expressed as follows 
\begin{equation}
 u(t,x,y)=u_1(t,y)+\cfrac{x}{t}, \qquad v(t,x,y)=v_1(t,y), \nonumber\end{equation}\begin{equation} p(t,x,y)=p_1(t,y), \qquad \rho(t,x,y)=\rho_1(t,y).
\end{equation}
The reduced equations read:
\begin{equation}
 \begin{cases}
\pd{\rho_1}{t}+\cfrac{\rho_1}{t}+\pd{\rho_1 v_1}{y}=0 
\\[5pt]
\rho_1\left(\pd{u_1}{t}+\cfrac{u_1}{t}+v_1\pd{u_1}{y}\right) = 
\mu\left(\pd{^2u_1}{y^2}\right)
\\[10pt]
\rho_1\left(\pd{v_1}{t}+v_1\pd{v_1}{y}\right) +\pd{p_1}{y}= 
\cfrac{4\mu}{3}\ \pd{^2v_1}{y^2}
\\[10pt]
\cfrac{C_v}{R}\left(\pd{p_1}{t}+\cfrac{p_1}{t}+\pd{p_1v_1}{y}\right)=\sigma:\tsr{S}+\cfrac{\kappa }{R}
\left(\pd{^2p_1/\rho_1}{y^2}\right)
\end{cases}\label{eqx5}
\end{equation}
with
\begin{equation}
 \cfrac{\sigma:\tsr{S}}{\mu}=-\cfrac{p_1}{\mu}\left(\cfrac{1}{t}+\pd{v_1}{y} \right)+\left(\pd{u_1}{y}\right)^2 
+\cfrac{4}{3}\left(\pd{v_1}{y}\right)^2+\cfrac{4}{3t^2} -\cfrac{4}{3t}\ \pd{v_1}{y} .
\end{equation}
The symmetries of (\ref{eqx5}) are generated by
\begin{multiequation} W_1=\pd{}{y}, \qquad W_2=\cfrac{1}{t}\pd{}{u_1}, \qquad W_3=t\pd{}{y}+\pd{}{v_1}, \\\\
 W_{4}=2t\pd{}{t}+y\pd{}{y}-u_1\pd{}{u_1}-v_1\pd{}{v_1}-2p_1\pd{}{p_1}, \\\\ W_5=y\pd{}{y}+u_1\pd{}{u_1}+v_1\pd{}{v_1}-2\rho_1\pd{}{\rho_1}. \end{multiequation}

\begin{figure}[ht]
	\centering
	\includegraphics[width=4cm]{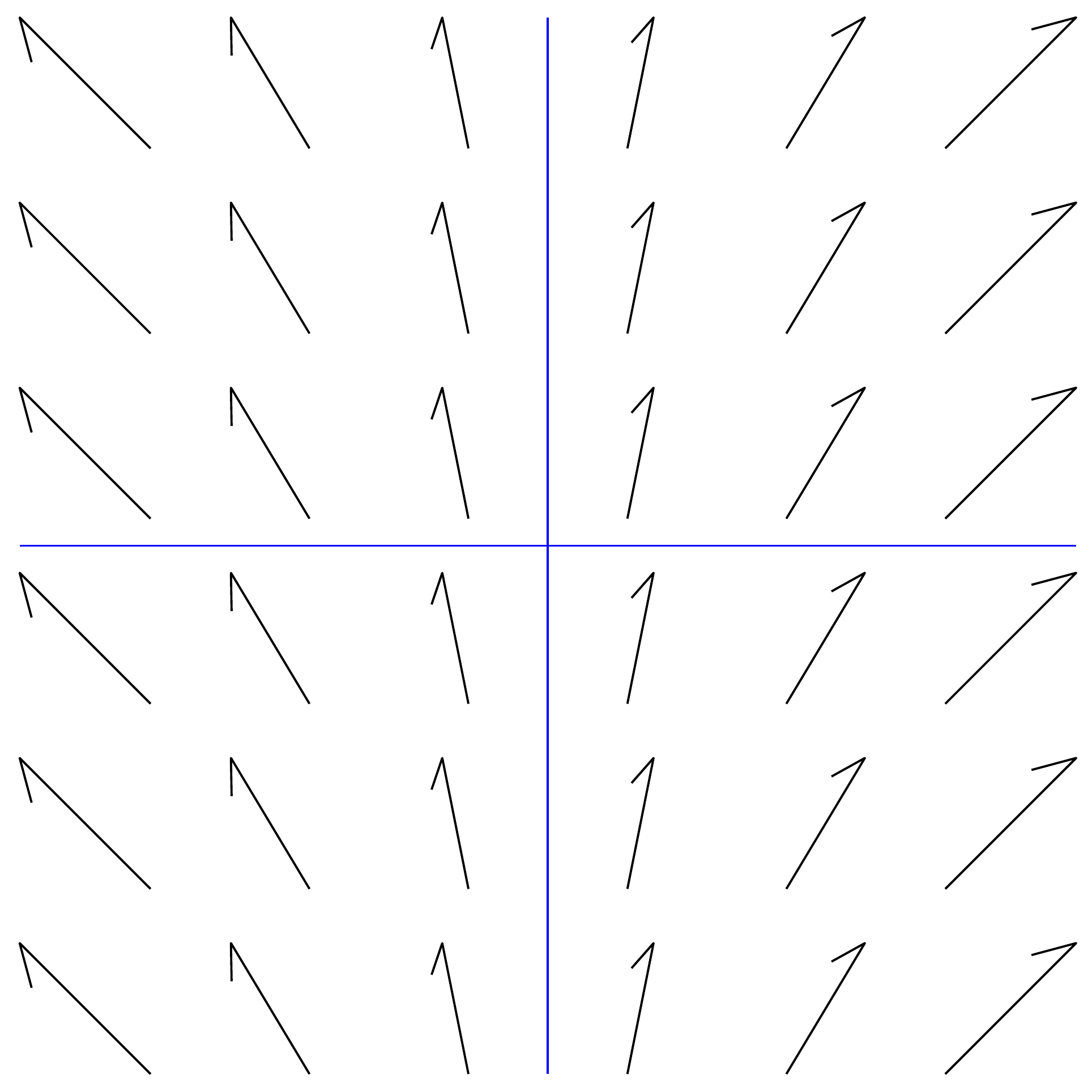}
	\caption{Solution (\ref{eq:unsteady_2d_x5y1}) with $t>0$ and $u_2=0$}
	\label{fig:unsteady_2d_x5y1}
\end{figure}

Solutions of (\ref{eqx5}) which are invariant under $W_1$ are $y$-independant. Solving equations (\ref{eqx5}) with this constraint leads to
\begin{equation}
 u_1(t,y)=\cfrac{u_2}{t}, \qquad v_1(t,y)=v_2,\qquad p_1(t,y)=p_2t^{(-1-R/C_v)}+\cfrac{4\mu}{3t}, \nonumber\end{equation}\begin{equation}  \rho_1(t,y)=\cfrac{\rho_2}{t}.
\end{equation}
where $u_2$, $v_2$, $p_2$ and $\rho_2$ are constants.  Consequently,
\begin{equation}
 \begin{mybox}
u(t,x,y)=\cfrac{x+u_2}{t}, \hspc  
v(t,x,y)=v_2, \\[5pt]
p(t,x,y)=p_2t^{(-1-R/C_v)}+\cfrac{4\mu}{3t}, \hspc\rho(t,x,y)=\cfrac{\rho_2}{t}.
\\[-10pt]\end{mybox}\label{eq:unsteady_2d_x5y1}\end{equation}
This solution is a rotation of solution (\ref{eq:unsteady_2d_y5}). It is presented in Figure \ref{fig:unsteady_2d_x5y1}.

\begin{figure}[ht]
	\centering
	\includegraphics[width=4cm]{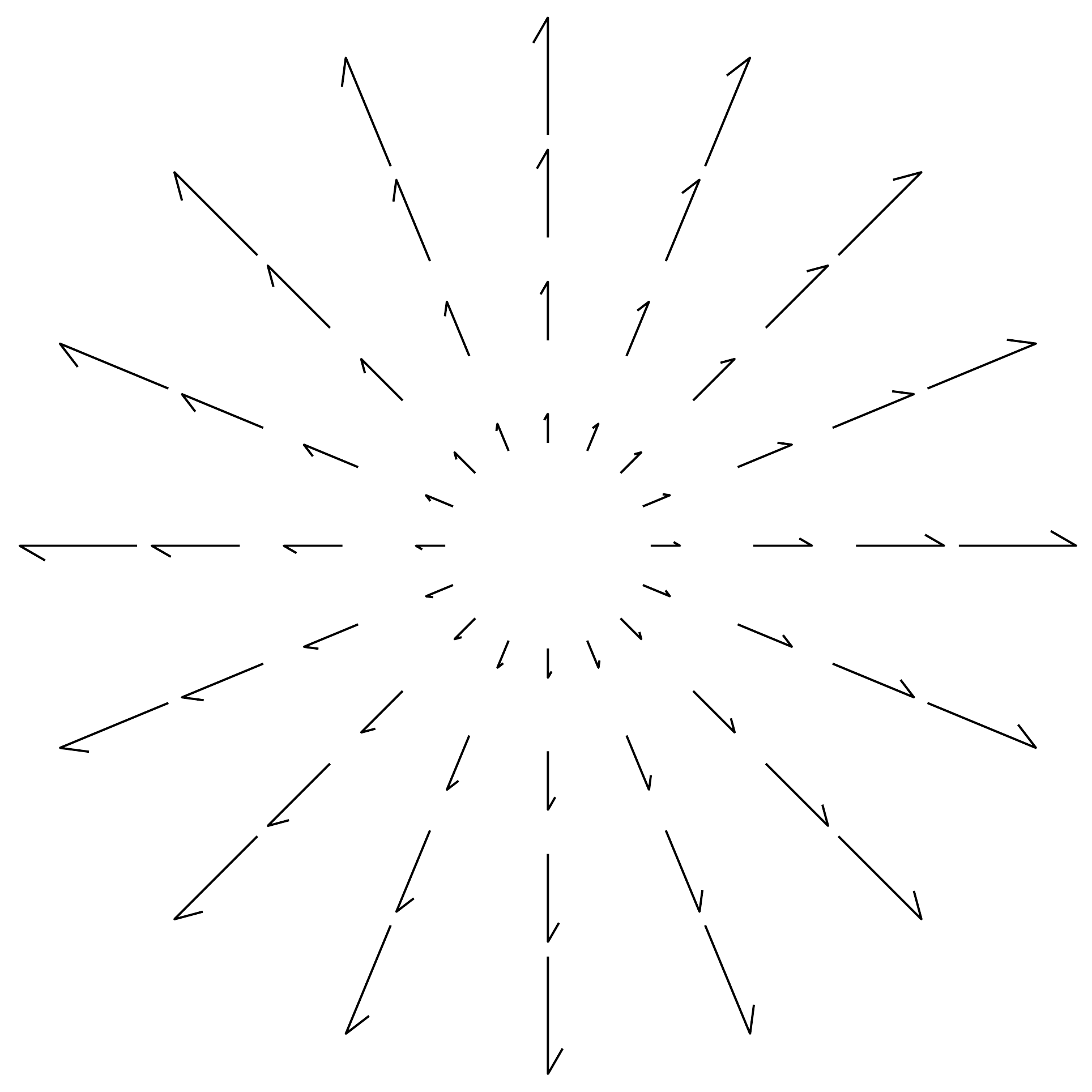}
	\caption{Solution (\ref{eq:unsteady_2d_x5y3}) with $t>0$}
	\label{fig:unsteady_2d_x5y3}
\end{figure}

If instead of $W_1$, we consider $W_3$ then the self-similar solutions of equations (\ref{eqx5}) are of the form:
\[ u_1(t,y)=u_2(t), \quad v_1(t,y)=v_2(t)+÷yt,\quad p_1(t,y)=p_2(t),\quad \rho_1(t,y)=ρ_2(t).\]
With these relations, the solutions to equations (\ref{eqx5}) is
\[ u_2=÷{u_3}t,\quad v_2=÷{v_3}t,\quad ρ_2=÷{ρ_3}t,\quad p_2=÷{4Rµ}{3(2R+C_v)t}+p_3t^{-2-2R/C_v}\]
where $u_3,v_3,p_3$ and $ρ_3$ are arbitrary scalars. The corresponding solution of (\ref{nsc}) is
\begin{equation}
 \begin{mybox}
	 u(t,x,y)=\cfrac{x+u_3}{t}, \qquad\qquad
v(t,x,y)=\cfrac{v_3+y}{t}, \\[5pt]\displaystyle 
p(t,x,y)=÷{4Rµ}{3(2R+C_v)t}+p_3t^{-2-2R/C_v}, \hspc\rho(t,x,y)=\cfrac{\rho_3}{t}.
\\[-10pt]\end{mybox}\label{eq:unsteady_2d_x5y3}\end{equation}
It represents a source flow around the point $(-u_3,-v_3)$. The velocity field is plotted in Figure \ref{fig:unsteady_2d_x5y3}.

\begin{figure}[t]
	\centering
	\includegraphics[width=37mm]{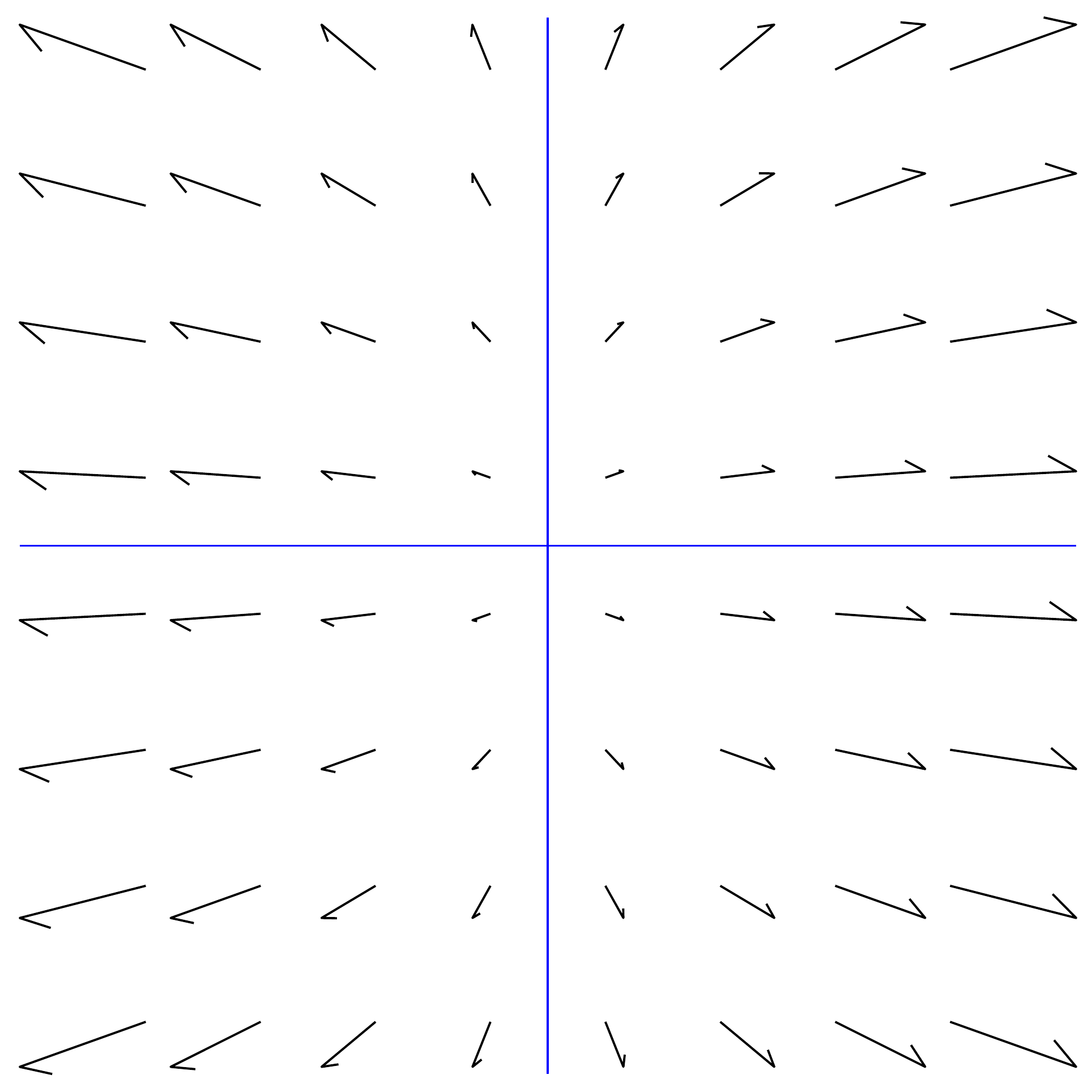}\hfill
	\includegraphics[width=37mm]{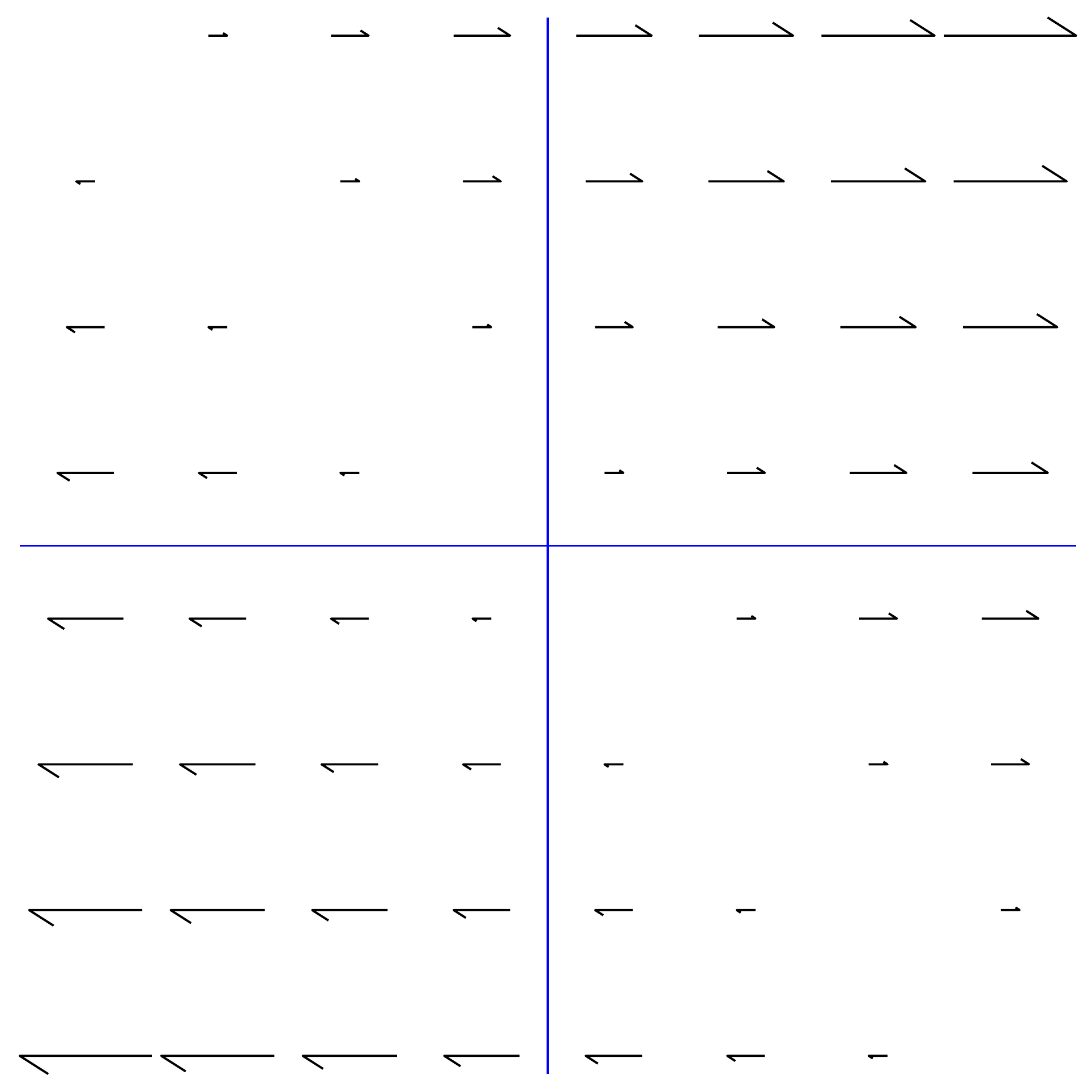}\hfill
	\includegraphics[width=37mm]{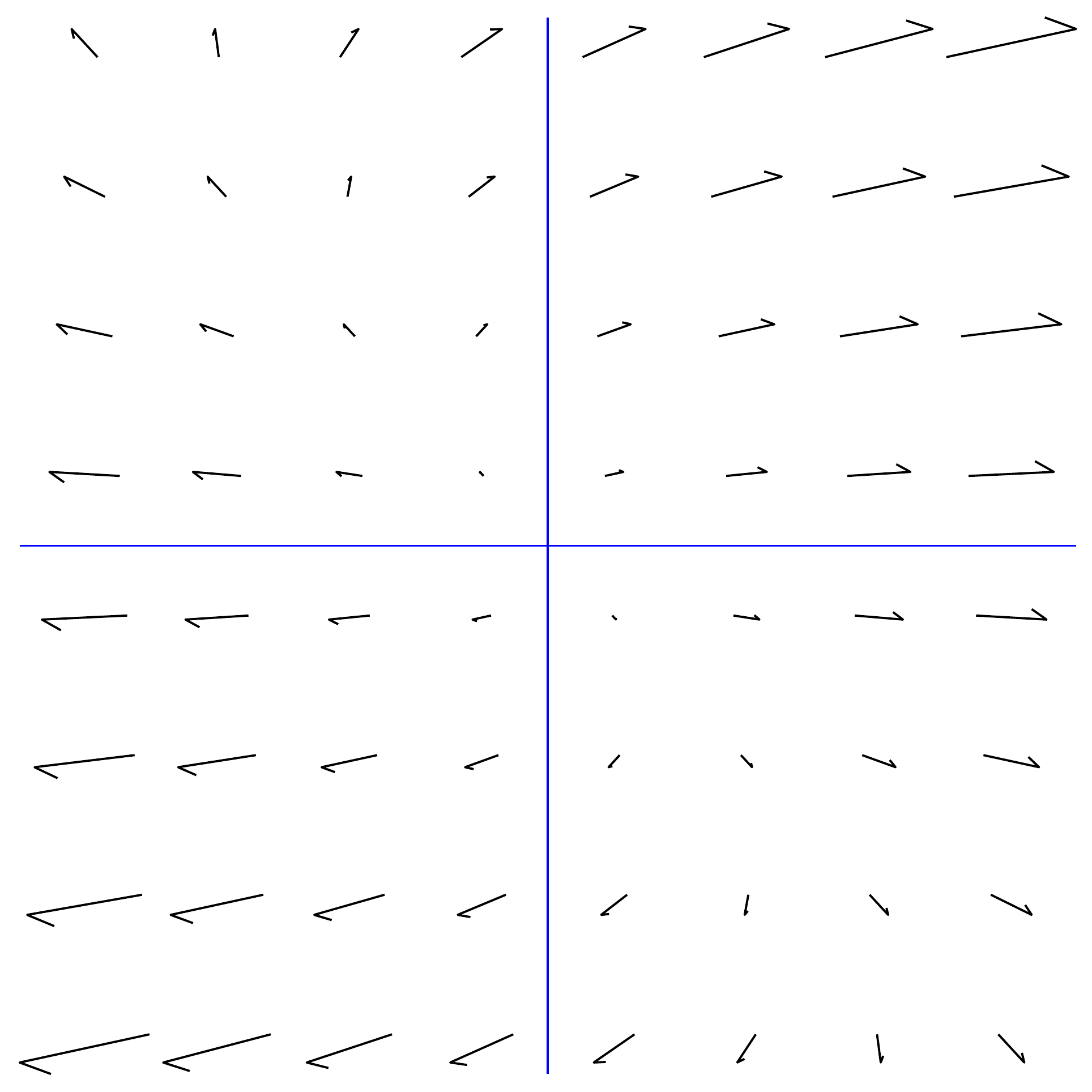}
	\caption{Solutions (\ref{eq:unsteady_2d_x5y5}) with $t>0$ and $v_3=0$. \\Left: $u_3=0,δ=1$. Center: $u_3>0,δ=0$. Right: $u_3>0,δ=1$}
	\label{fig:unsteady_2d_x5y5}
\end{figure}

To find self-similar solutions under infinitesimal scale transformation $W_5$, we set:
 \begin{multiequation}
 u_1(t,y)=u_2(t)y, \qquad v_1(t,y)=v_2(t)y, \\\\ p_1(t,y)=p_2(t), \qquad\rho_1(t,y)=\cfrac{\rho_2(t)}{y^2}.
 \end{multiequation}
It follows from the reduced equations that
\begin{equation}
 v_2(t)=\cfrac{\delta}{t+v_3}, \qquad u_2(t)=\cfrac{u_3}{t(t+v_3)^\delta}, \qquad \rho_2(t)=\cfrac{(t+v_3)^\delta\rho_3}{t}
\end{equation}
where $u_3$ and $\rho_3$ are constants and $\delta=0$ or $1$. Thus,
\begin{equation}
	\begin{mybox}\\[-10pt]
u(t,x,y)=\cfrac{u_3y}{t(t+v_3)^\delta}+\cfrac{x}{t}, \quad v(t,x,y)=\cfrac{y\delta }{t+v_3}, \quad
\rho(t,x,y)=\cfrac{(t+v_3)^\delta\rho_3}{ty^2}.\\[-10pt]\end{mybox}\label{eq:unsteady_2d_x5y5}
\end{equation}
The pressure $p(t)$ is the solution of the ordinary differential equation:
\begin{equation}
\cfrac{C_v}{R}p' +\cfrac{(C_v+R)(2t+v_3)^\delta\rho_3-2\kappa t^2}{\rho_3Rt(t+v_3)^\delta}\ p=\mu\cfrac{3u_3^2+4(t^2+v_3t+v_3^2)^\delta}{3t^2(t+v_3)^{2\delta}}.
\end{equation}
If $u_3=0$ and $δ=0$, the velocity field is similar to Figure \ref{unsteady_2d_y5} but 90 degrees rotated.
In other cases, the flow is graphically presented in Figure \ref{fig:unsteady_2d_x5y5}.

Another solution of equations (\ref{eqx5}) can be found by setting $v_1$ constant and $u_1$ linear in $y$. This leads to the following solution of (\ref{nsc})
\begin{equation}
 \begin{mybox}
u(t,x,y)=\cfrac{x+y+u_2}{t}-v_2, \hspc  
v(t,x,y)=v_2, \\[5pt]
\rho(t,x,y)=\cfrac{1}{t}\ \cfrac{1}{ (y-v_2t)^2\rho_3+(y-v_2t)\rho_4+\rho_5}
\\[-10pt]\end{mybox}\end{equation}
where $u_2$, $v_2$, $\rho_3$, $\rho_4$ and $\rho_5$ are constants. $p(t)$ is the solution of
\begin{equation}
 \cfrac{C_v}{R}p'+\cfrac{C_v+R-2\kappa \rho_3t^2}{Rt}\ p-\cfrac{7\mu}{t^2}=0.\label{eq:unsteady_2d_x5}
\end{equation}
\begin{figure}[ht]
	\centering
	\includegraphics[width=4cm]{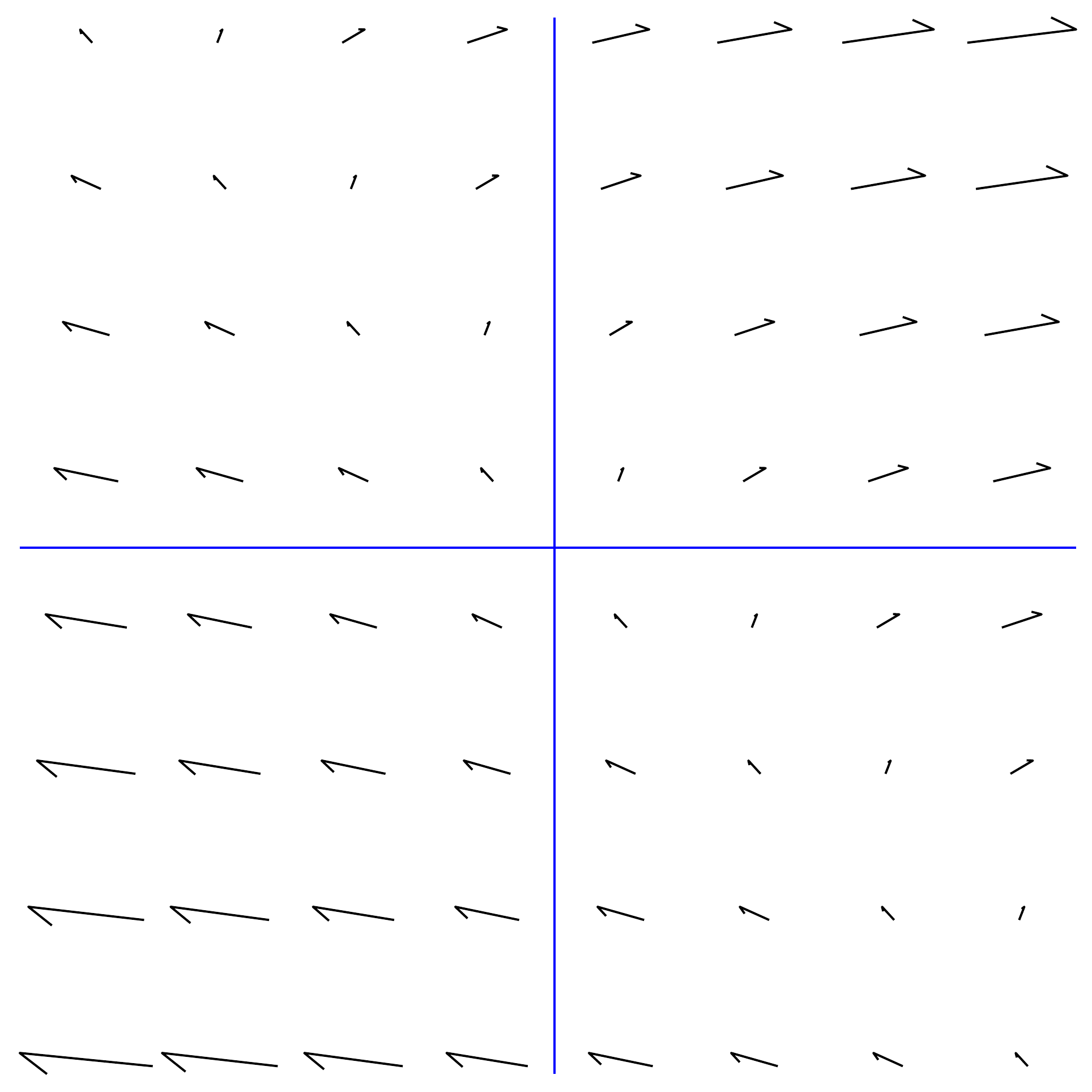}
	\caption{Solution (\ref{eq:unsteady_2d_x5}) with $t=1$ and $v_2=1$ and $u_2=0$}
	\label{fig:unsteady_2d_x5}
\end{figure}

In the next subsection, we calculate bidimensional solutions of (\ref{nsc}) which are self-similar under scale transformations.

\subsection{Scale transformations}

Infinitesimal symmetry $X_{11}$ suggests a change of variables:
\begin{multiequation}
 u(t,x,y)=\cfrac{u_1(\chi,\eta)}{\sqrt{t}}, \qquad v(t,x,y)=\cfrac{v_1(\chi,\eta)}{\sqrt{t}}, \qquad p(t,x,y)=\cfrac{p_1(\chi,\eta)}{t}, \\
\rho(t,x,y)=\rho_1(\chi,\eta)  \eqspc{with}\chi=\cfrac{x}{\sqrt{t}}, \qquad \eta=\cfrac{y}{\sqrt{t}}.
\end{multiequation}
The equations of the new variables are:
\[\begin{cases}
 -\cfrac{\chi}{2}\pd{\rho_1}{\chi}-\cfrac{\eta}{2}\pd{\rho_1}{\eta}+\pd{\rho_1u_1}{\chi}+\pd{\rho_1v_1}{\eta}=0
\\\\
\rho_1\left(-\cfrac{\chi}{2}\pd{u_1}{\chi}-\cfrac{\eta}{2}\pd{u_1}{\eta}-\cfrac{1}{2}u_1+u_1\pd{u_1}{\chi} +v_1\pd{u_1}{\eta}\right) =
\\\hspace{7.5em}-\pd{p_1}{\chi}+\cfrac{\mu}{3}\left(4\pd{^2u_1}{χ^2}+\ppd{v_1}{\chi}{\eta}+3\pd{^2u_1}{η^2}\right)
\\[10pt]
\rho_1\left(-\cfrac{\chi}{2}\pd{v_1}{\chi}-\cfrac{\eta}{2}\pd{v_1}{\eta}-\cfrac{1}{2}v_1+u_1\pd{v_1}{\chi} +v_1\pd{v_1}{\eta}\right) =
\\\hspace{7.5em}-\pd{p_1}{\eta}+\cfrac{\mu}{3}\left(4\pd{^2v_1}{η^2}+\ppd{u_1}{\chi}{\eta}+3\pd{^2v_1}{χ^2}\right)
\\[10pt]
\cfrac{C_v}{R}\left(-\cfrac{\chi}{2}\pd{p_1}{\chi}-\cfrac{\eta}{2}\pd{p_1}{\eta}-p_1+\pd{p_1u_1}{\chi} +\pd{p_1v_1}{\eta}\right) =
\\\hspace{7.5em}-p⦅\pd {u_1}{χ}+\pd {v_1}{η}⦆+µS'+\cfrac{\kappa}{R} ⦅\pd{^2}{χ^2}+\pd{^2}{η^2}⦆\cfrac{p_1}{\rho_1}
\end{cases}\]
where 
\[
S'=÷43⟦⦅\pd {u_1}{χ}⦆^2+⦅\pd {v_1}{η}⦆^2-\pd {u_1}{χ}\pd {v_1}{η}⟧+⦅\pd {u_1}{η}+\pd {v_1}{χ}⦆^2.
\]
A solution to these equations is
\begin{equation}
	\begin{array}{l}
		u_1(\chi,\eta)=u_2\chi, \qquad v_1(\chi,\eta)=0, \qquad \rho_1(\chi,\eta)=\cfrac{\rho_2}{\chi^2}, \\ p_1(\chi,\eta)=\cfrac{4\mu^2\rho_2R}{3\rho_2R-6\kappa}
	\end{array}
\end{equation}
for some constants $u_2$ and $ρ_2$.  We deduce that
\begin{equation}
 \begin{mybox}
u(t,x,y)=\cfrac{u_2x}{t}, \hspc
v(t,x,y)=0, \\[10pt]
 p(t,x,y)=\cfrac{4\mu^2\rho_2R}{3(\rho_2R-2\kappa)t}, \hspc \rho(t,x,y)=\cfrac{\rho_2t}{x^2}.
\\[-10pt]\end{mybox}\end{equation}


To get self-similar solutions of the bidimension equations (\ref{unsteady_bidim}) under $X_{12}$, one makes the change of variables:
\begin{eqnarray}\eta=\cfrac{y}{x}, \qquad u_1(t,\eta)=\cfrac{u(t,x,y)}{x}, \qquad v_1(t,\eta)=\cfrac{v(t,x,y)}{x}, \\[10pt] p_1(t,\eta)=p(t,x,y), \qquad \rho_1(t,\eta)=x^2\rho(t,x,y).\end{eqnarray}
The equations become:
\begin{equation}\begin{cases}
		\pd{\rho_1}{t}+\pd {}{η}\bigg[\rho_1(v_1-u_1\eta)\bigg]=0 \\[10pt]
		\rho_1\left[\pd{u_1}{t}+u_1^2+(v_1-u_1\eta) \pd{u_1}{η}\right]-\eta\pd {p_1}{η} =\mu\left(\cfrac{4\eta^2}{3}+1\right) \ppd{u_1}{η}{η}-\mu\cfrac{η}{3} \ppd{v_1}{η}{η}
\\[10pt]
\rho_1\left[\pd{v_1}{t}+u_1v_1+(v_1-u_1η)\pd{v_1}{η}\right]+\pd {p_1}{η}=-µ\cfrac{η}{3}  \ppd{u_1}{η}{η}+ \mu  \left(\cfrac{4}{3}+\eta^2\right) \ppd{v_1}{η}{η}
\\[10pt]\displaystyle 
\cfrac{C_v}{R}\left[\pd{p_1}{t}+(v_1-ηu_1)\pd {p_1}{\eta}\right]= -⦅÷{C_v}{R}+1⦆p_1⦅u_1-η\pd {u_1}{η}+\pd {v_1}{η}⦆\\
\hspace{35mm} +µ\tsr D+\cfrac{\kappa}{R}\left[(1+\eta^2)\ppd{}{η}{η}\cfrac{p_1}{\rho_1} -2\eta\pd{}{η}\cfrac{p_1}{\rho_1}+2\cfrac{p_1}{\rho_1}\right]

\end{cases}\label{eqx2x12}\end{equation}
where
\[
\begin{array}{r}\displaystyle 
	\tsr D=-÷23⦅u_1-η\pd {u_1}{η}+\pd {v_1}{η}⦆^2+2⦅u_1-η\pd {u_1}{η}⦆^2+2⦅\pd {v_1}{η}⦆^2
	\\\displaystyle 
	+⦅\pd {u_1}{η}-η\pd {v_1}{η}+v_1⦆^2
\end{array}
\]
A solution of (\ref{eqx2x12}) is
\be u_1(t,\eta)=\cfrac{1}{t}, \qquad v_1(t,\eta)=\cfrac{\eta}{t}, \qquad p_1(t,\eta)=p_2(t), \qquad \rho_1(t,\eta)=\rho_2 \ee
where $\rho_2$ is a constant, and $p_2(t)$ is the solution of
\be 3C_v\rho_2t^2p'_2(t)+6t(C_v\rho_2-\kappa t+\rho_2R)p_2(t)-4\mu\rho_2R=0 \label{whitt}\ee
With the original variables, we get:
\begin{equation}\begin{mybox}u(t,x,y)=\cfrac{x}{t}, \hspc v(t,x,y)=\cfrac{y}{t}, \\ p(t,x,y)=p_2(t), \hspc \rho(t,x,y)=\cfrac{\rho_2}{x^2}.\\[-10pt]\end{mybox}\end{equation}
The pressure $p_2(t)$ can be written in terms of Whittaker $M$ functions \cite{olver10}. For instance, when $C_v\rho_2=1$, if we call $b=\rho_2R$ then
\be p_2(t)=\frac{2b\mu \e^{\kappa t} M_{b,b+1/2}(2\kappa t)}
{3(2b+1)\kappa t^2(2\kappa t)^b}+p_3\cfrac{e^{2\kappa t}}{t^{2+2b}}
\ee
$p_3$ being a constant.

Other solutions of (\ref{eqx2x12}) can be obtained knowing that these equations admit the following infinitesimal symmetry:
$$(1+\eta^2)\pd{}{\eta}+(\eta u_1-v_1)\pd{}{u_1}+(\eta v_1+u_1)\pd{}{v_1}-2\rho_1\eta\pd{}{\rho_1}.$$
It suggests the change of variables:
\begin{equation}
	\begin{array}{l}
		u_1(t,\eta)=u_2(t)\eta+v_2(t), \qquad v_1(t,\eta)=v_2(t)\eta-u_2(t), \\[5pt] p_1(t,\eta)=p_2(t), \qquad \rho_1(t,\eta)=\cfrac{\rho_2(t)}{1+\eta^2}, 
	\end{array}
\label{vareqx2x12y3}\end{equation}
corresponding to a velocity field
\[
\vt u(t,r,θ)=v_2(t)r\vt e_r+u_2(t)r\vt e_{θ}.
\]

\begin{figure}[ht]
	\centering
	\includegraphics[width=4cm]{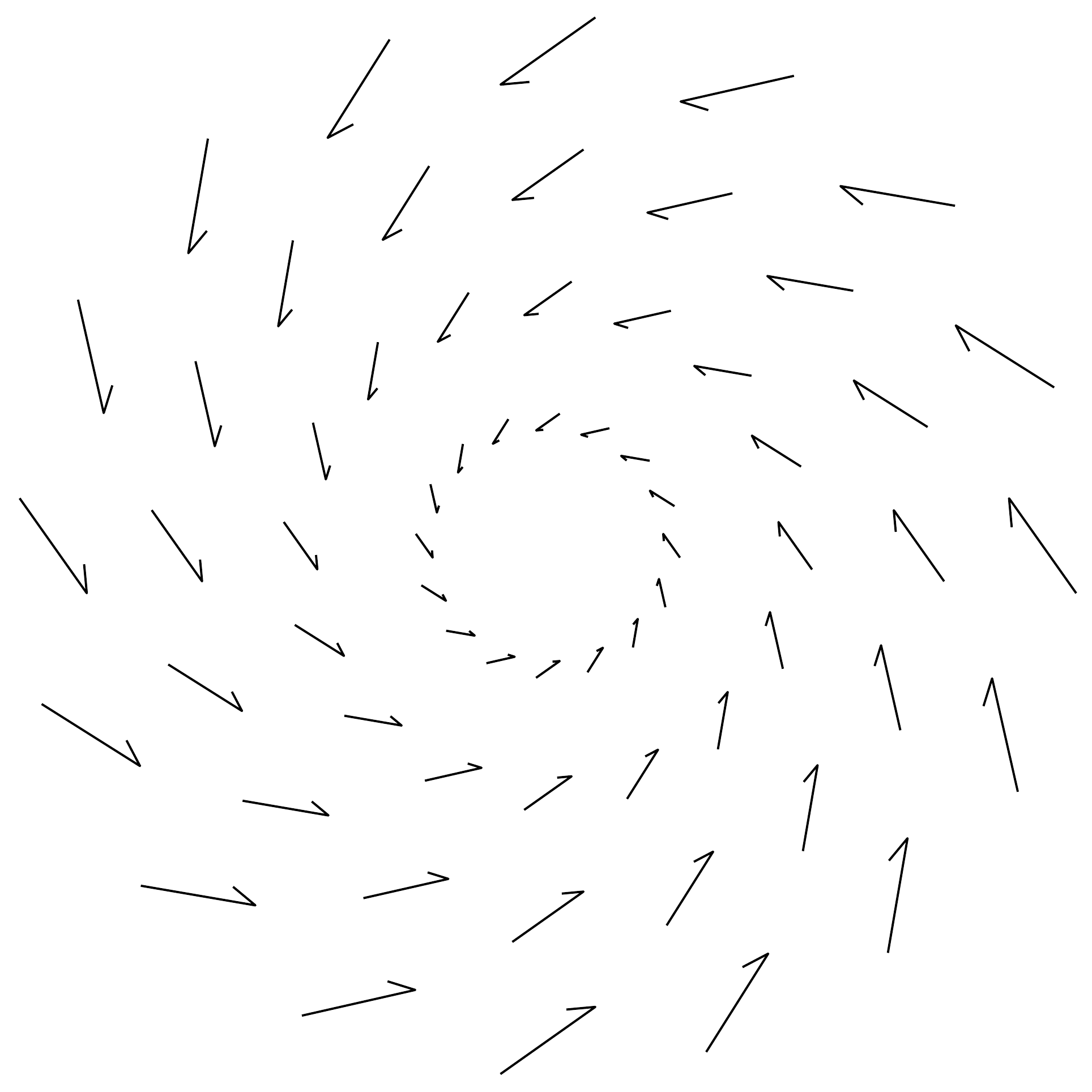}\qquad
	\includegraphics[width=4cm]{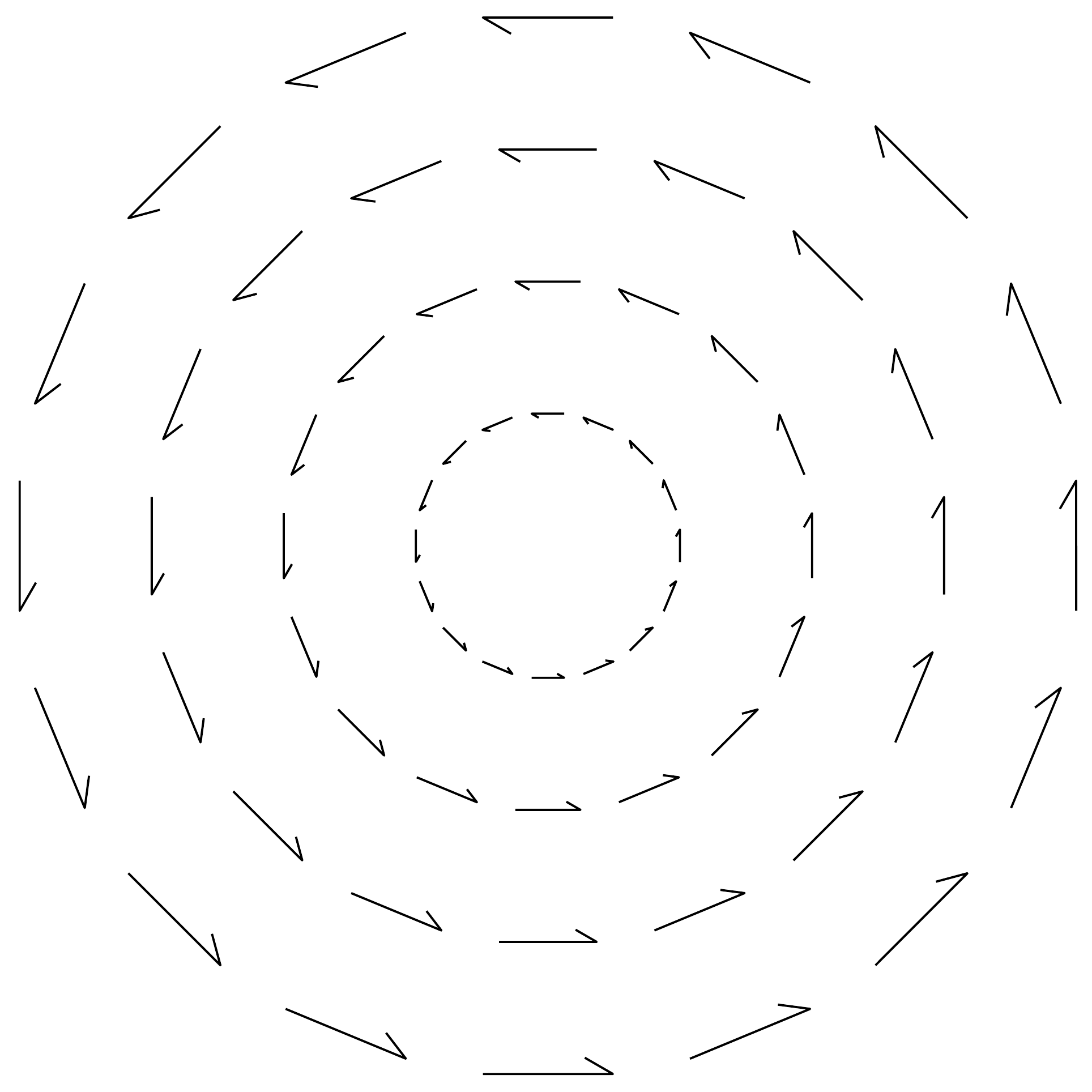}

	\vspace{\baselineskip}
	\includegraphics[width=4cm]{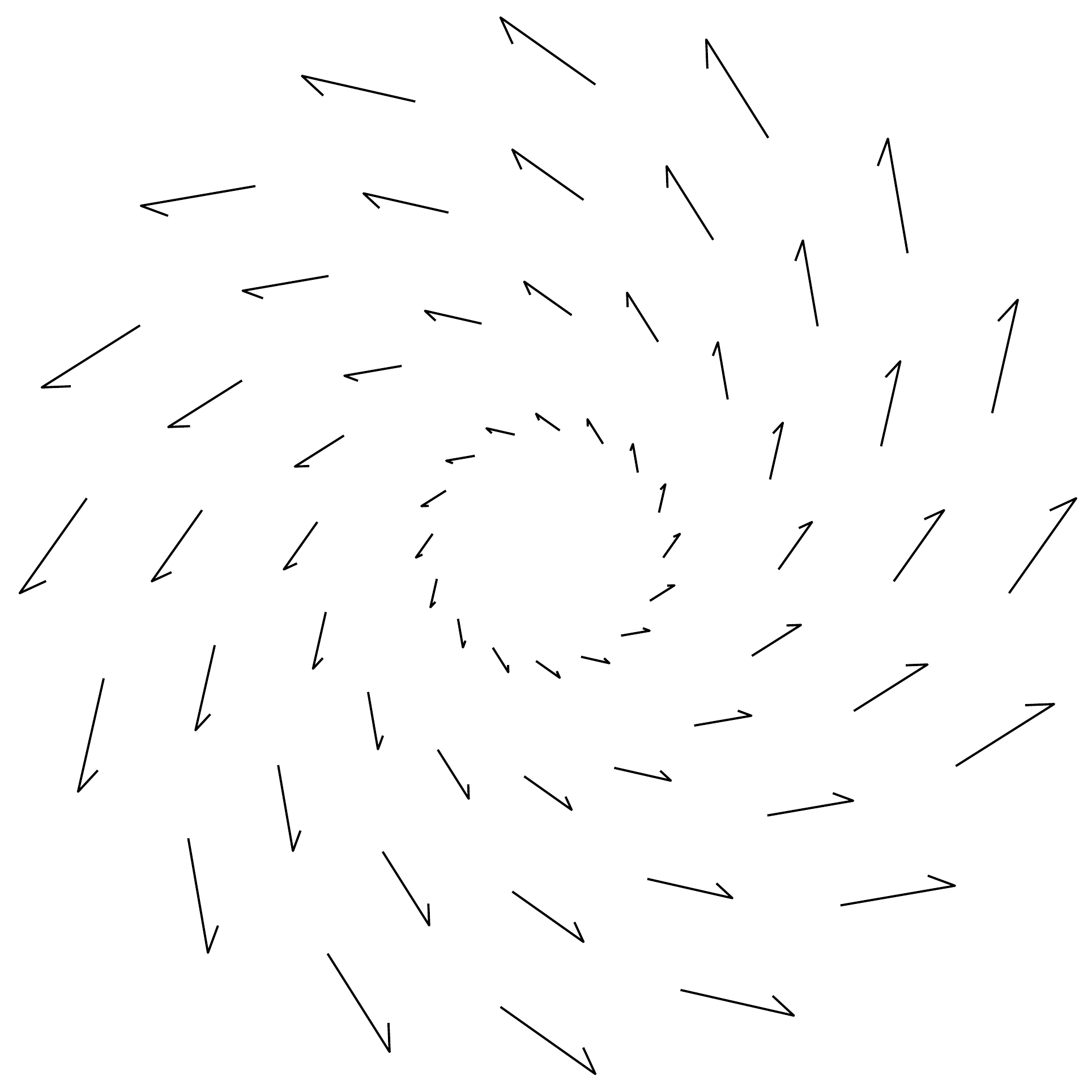}\qquad
	\includegraphics[width=4cm]{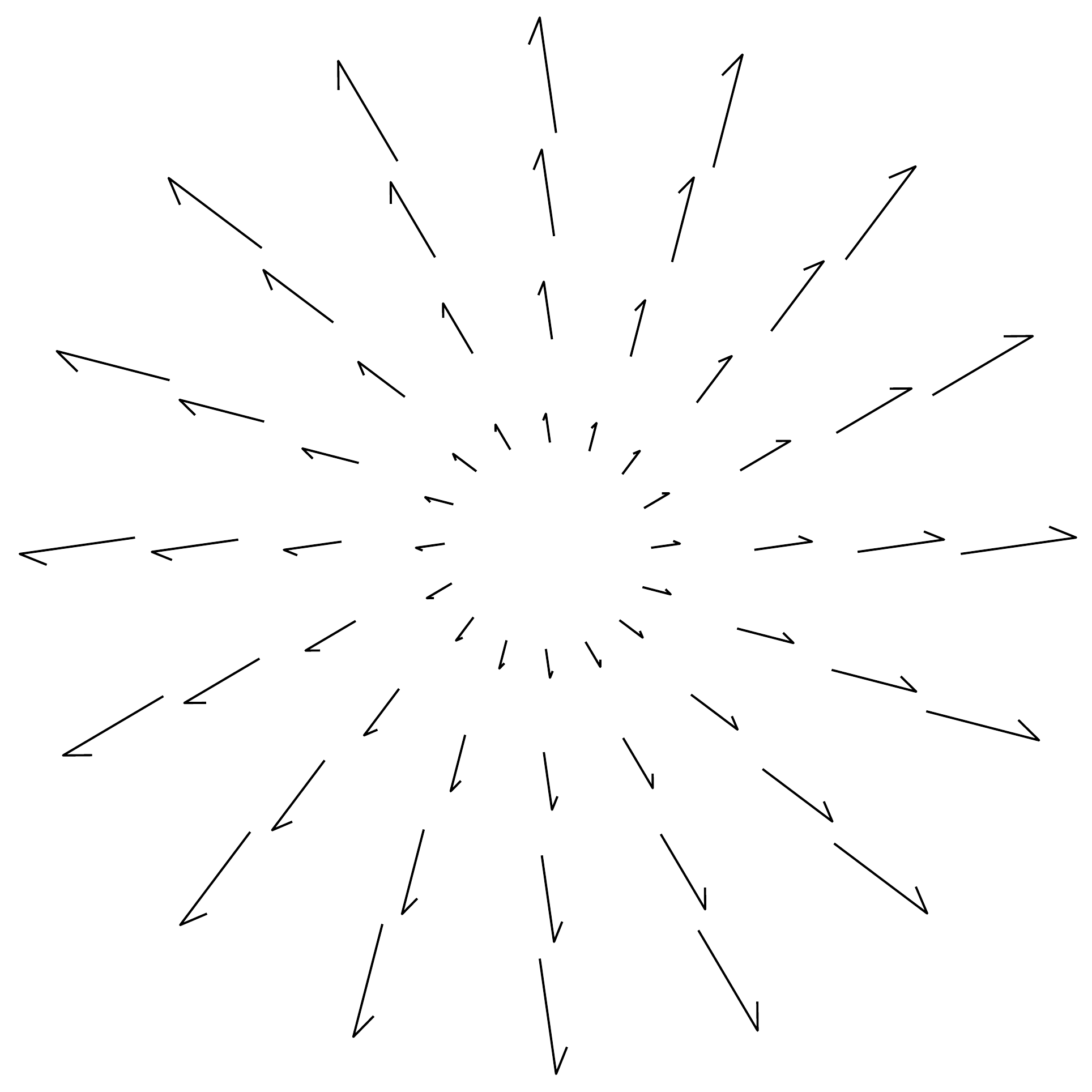}
	\caption{Solution (\ref{eq:unsteady_2d_x5z}) with $b=0$ and $a=1$, at $t=-1$ (top left), $t=0$ (top right), $t=1$ (bottom left) and $t=10$ (bottom right)}
	\label{fig:unsteady_2d_x5z}
\end{figure}

Inserting these relations into equations (\ref{eqx2x12}), we obtain:
\[
v_2(t)=÷{f'(t)}{2f(t)},\qquad u_2(t)=±√{v_2^2(t)+v_2'(t)},\qquad  ρ_2(t)=ρ_3
\]
\[
p_2(t)=÷{µR}3\,÷{h(t)}{f(t)^{1+R/C_v}}\ ∫÷{f(t)^{R/C_v-1}f'(t)^2}{h(t)}ｄt
\]
with $f(t)=at^2+2bt+2$, $a$, $b$ and $ρ_3$ being constants such that $2a\geq b^2$. It follows that
\begin{equation}
	\begin{mybox}
		\\[-10pt]\displaystyle
		\u(t,r,θ)=÷{at+b}{at^2+2bt+2}r\ve_r±÷{√{2a-b^2}}{at^2+2bt+2}r\ve_{θ}, \qquad ρ(t,r,θ)=÷{ρ_3}{r^2},
		\\[10pt]\displaystyle 
		p(t,r,θ)=÷{4µR\e^{4κt/C_vρ_3}}{(at^2+2bt+2)^{R/C_v+1}} ∫÷{(at^2+2bt+2)^{R/C_v-1}(at+b)^2}{3C_v\e^{4κt/C_vρ_3}}ｄt.
		\\[-10pt]
	\end{mybox}
	\label{eq:unsteady_2d_x5z}
\end{equation}
If $a=b^2/2$ then the flow is radial, with $\vt u=\cfrac r{t+2/b}\vt e_r$. The velocity field is similar to that of solution (\ref{eq:unsteady_2d_x5y3}), plotted in Figure \ref{fig:unsteady_2d_x5y3}, for a fixed $t$ but the pressure and density fields are different. The evolution of the flow with time can be visualized in Figure \ref{fig:unsteady_2d_x5z} in the case $b=0$, $a=1$ and $u_2(t)>0$.

In the following section, we compute some three-dimensional solutions of equations (\ref{nsc}).

\section{3D-solutions}\label{3d}

We first consider solutions invariant under $X_1$, $X_2$ and $X_4$. Such solutions depend only on $y$:
\begin{eqnarray}u(t,x,y,z)=u_1(y), \qquad v(t,x,y,z)=v_1(y), \qquad w(t,x,y,z)=w_1(y), \\\\
p(t,x,y,z)=p_1(y), \qquad \rho(t,x,y,z)=\rho_1(y).\end{eqnarray}
The reduced equations are:
\begin{equation}\begin{cases}
 (\rho_1v_1)'=0 ,\\
 \rho_1v_1u_1'=\mu u_1'',\\
 3\rho_1v_1v'_1+p_1'=4\mu v_1'',\\
 \rho_1v_1w_1'=\mu w_1'',\\
 3C_v(p_1v_1)'=R(4\mu v_1'^2-3v_1'p_1+3\mu u_1'^2+3\mu w_1'^2)+3\kappa (p_2/\rho_2)''.
\end{cases}\end{equation}
We find the following solution which is an extension of (\ref{solx1x2x4}):
\begin{equation}\begin{mybox}\\[-10pt]\displaystyle 
		u(t,y)=u_3\pm\sqrt{\frac{(2C_v\mu-4\kappa-3R\mu)v_3^2-3R\mu w_3^2}{3R\mu}}\ \e^{ay/µ}, 
		\\[15pt] v(t,y)=v_3\e^{ay/µ},\qquad w(t,x,y,z)=w_3\e^{ay/\mu}, 
		\\[5pt]
 p(t,y)=\cfrac{av_3}{3}\e^{ay/µ}, \qquad \rho(t,y)=\cfrac{a}{v_3}\e^{-ay/µ}, \\[-10pt]
\end{mybox}\end{equation}
In these expressions, $a$, $u_3$ $v_3$ and $w_3$ are arbitrary constants.

The generator $(X_2+X_5)+(X_3+X_6)$ leads to the ansatz:
\begin{equation}
	\begin{array}{l}u(t,x,y,z)=u_1(t,z)+\cfrac{xu_0}{tu_0+a}, \qquad v(t,x,y,z)=v_1(t,z)+\cfrac{yv_0}{tv_0+b}, \\[10pt] w(t,x,y,z)=w_1(t,z), \quad p(t,x,y,z)=p_1(t,z), \quad \rho(t,x,y,z)=\rho_1(t,z)
	\end{array}
\end{equation}
where $a$, $b$, $u_0$ and $v_0$ are arbitrary constants. Equations (\ref{nsc}) reduce into:
\begin{equation}\begin{cases} \pd{\rho_1}{t}+w_1\pd{\rho_1}{z}+\rho_1δ=0,
		\\[10pt]
  \rho_1\left(\pd{u_1}{t}+\cfrac{u_0u_1}{u_0t+a}+w_1\pd{u_1}{z}\right)=\mu\pd{^2u}{z^2} ,
  \\[10pt]
  \rho_1\left(\pd{v_1}{t}+\cfrac{v_0v_1}{v_0t+b}+w_1\pd{v_1}{z}\right)=\mu\pd{^2v}{z^2} ,
  \\[10pt]
\rho_1\left(\pd{w_1}{t}+w_1\pd{w_1}{z}\right)+\pd{p_1}{z}=\cfrac{4\mu}{3}\ \pd{^2w}{z^2} ,
\\[10pt]
\cfrac{C_v}{R}\left(\pd{p_1}{t}+w_1\pd{p_1}{z}\right)=⦅\cfrac{C_v}R+1⦆δ p_1+µ\tsr S_2+\cfrac{\kappa}{R}\pd{^2}{z^2}⦅\cfrac{p_1}{\rho_1}⦆
  \end{cases}
\label{eqx5x6}\end{equation}
where 
\[
δ=\cfrac{u_0}{u_0t+a}+\cfrac{v_0}{v_0t+b}+\pd{w_1}{z}
\]
is the divergence of $\u$ and
\[
\tsr S_2=-÷23δ^2+÷{2u_0^2}{(tu_0+a)^2}+÷{2v_0^2}{(tv_0+b)^2}+⦅\pd {u_1}z⦆^2\!+⦅\pd {v_1}z⦆^2\!+2⦅\pd {w_1}z⦆^2\!.
\]
The infinitesimal symmetries of these equations are
\begin{equation}
	\begin{array}{l} 
		R_1=\pd{}{z}, \qquad R_2=t\pd{}{z}+\pd{}{w}, \qquad  R_3=\cfrac{1}{u_0t+a}\pd{}{u}, 
		\\[10pt] 
		R_4=\cfrac{1}{v_0t+b}\pd{}{v}, \qquad  R_5=z\pd{}{z}+u\pd{}{u}+v\pd{}{v}+w\pd{}{w}-2\rho\pd{}{\rho}.
	\end{array}
\end{equation}
Without loss of generalty, assume that $u_0=v_0=1$.

Symmetry $R_1$ reduces equations (\ref{eqx5x6}) into:
\begin{equation}\begin{cases}
 (t+a)(t+b)\rho_2'+\rho_2=0\\
 (t+a)u_2'+u_2=0\\
 (t+b)v_2'+v_2=0\\
 w_2'=0\\
 C_vp_2'=R\ \sigma:\tsr{S}
\end{cases}\end{equation}
where 
\begin{eqnarray} u_2(t)=u_1(t,z), \qquad v_2(t)=v_1(t,z), \qquad w_2(t)=w_1(t,z), \\[10pt] p_2(t)=p_1(t,z), \qquad\rho_2(t)=\rho_1(t,z). \end{eqnarray}
The resolution of the equations gives the bidimensional solution
\begin{equation}\begin{mybox}
 u(t,x,y,z)=\cfrac{x+u_3}{t+a}, \qquad v(t,x,y,z)=\cfrac{y+v_3}{t+b}, 
 \\[5pt]
 w(t,x,y,z)=w_3, \qquad 
 \rho(t,x,y,z)=\cfrac{\rho_3}{h(t)},
 \\[10pt]\displaystyle 
 p(t,x,y,z)=\cfrac{4R\mu}{3C_v}h(t)^{-R/C_v-1} 
 \int \cfrac{h(t)+(a-b)^2}{h(t)^{1-R/C_v}}\d t
\end{mybox}\label{3d_x2x5_x3x6_y1}\end{equation}
where $u_3$, $v_3$, $w_3$ and $ρ_3$ are arbirtary constants and
\[
h(t)=(t+a)(t+b).
\]
In the particular case where $a=b$, the expression of $p$ in (\ref{3d_x2x5_x3x6_y1}) simplifies into
\[ p(t,x,y,z)= \cfrac{4R\mu}{3(2R+C_v)t}+p_3t^{-2R/C_v-2},\]
$p_3$ being a constant.

An invariant solution of equations (\ref{eqx5x6}) under $R_2$ has the following form:
\begin{equation}
	\begin{array}{l} u_1(t,z)=u_2(t),\quad v_1(t,z)=v_2(t),\quad w_1(t,z)=w_2(t)+\cfrac{zw_0}{tw_0+c}
		\\[10pt]
		p1(t,z)=p_2(t),\quad ρ_1(t,z)=ρ_2(t).
	\end{array}
	\label{varx5x6r2}
\end{equation}
Assume that $w_0=1$. A straightforward solution of (\ref{eqx5x6}), when $a=0$, is
\be\begin{mybox}
u(t,x,y,z)=\cfrac{x+u_3}{t}, \qquad v(t,x,y,z)=\cfrac{y+v_3}{t+b}, 
\\[10pt]
w(t,x,y,z)=\cfrac{z+w_3}{t+c}, \qquad \rho(t,x,y,z)=\cfrac{\rho_3}{h(t)},\\[10pt]
p(t,x,y,z)=\cfrac{4R\mu}{3C_v}\ h(t)^{-1-\frac{R}{C_v}} \displaystyle\int ÷{t^2(b-c)^2+bc(t+b)(t+c)}{h(t)^{1-\frac{R}{C_v}}}\d t.
\\[-10pt]\end{mybox}
	\label{eqx5x6r2_1}
\ee
where $u_3$, $v_3$, $w_3$, $\rho_3$ are constants and
\begin{equation}
 h(t)=t(t+b)(t+c).
\end{equation}
Another solution of (\ref{eqx5x6}) in the form (\ref{varx5x6r2}), with $a=b=c=t_0$, is
\be\begin{mybox}
u(t,x,y,z)=\cfrac{x+u_3}{t+t_0}, \qquad v(t,x,y,z)=\cfrac{y+v_3}{t+t_0}, 
\\[10pt]
w(t,x,y,z)=\cfrac{z+w_3}{t+t_0}, \qquad \rho(t,x,y,z)=\cfrac{\rho_3}{(t+t_0)^3},\\[10pt]
p(t,x,y,z)=p_3(t+t_0)^{-3-\frac{3R}{C_v}} 
\\[-10pt]\end{mybox}
	\label{eqx5x6r2_2}
\ee
where $u_3$, $v_3$, $w_3$, $\rho_3$ are constants. In (\ref{eqx5x6r2_1}) and (\ref{eqx5x6r2_2}), the flow is fully three-dimensional. It is purely radial around the source point $(u_3,v_3,w_3)$. The pressure and the density are uniform but time-dependent.

\begin{figure}[ht]\centering
\includegraphics[width=5cm]{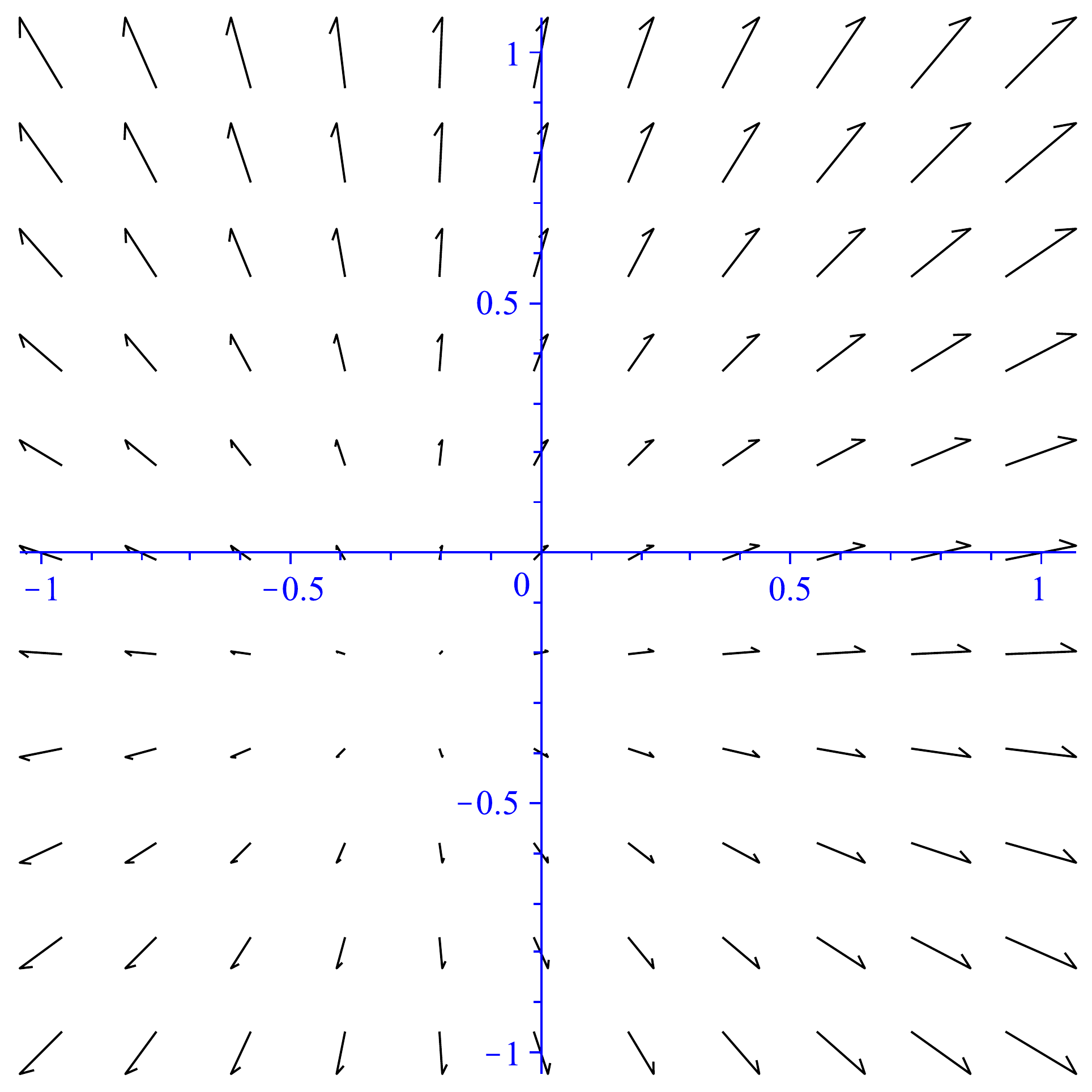}\qquad 
\includegraphics[width=5cm]{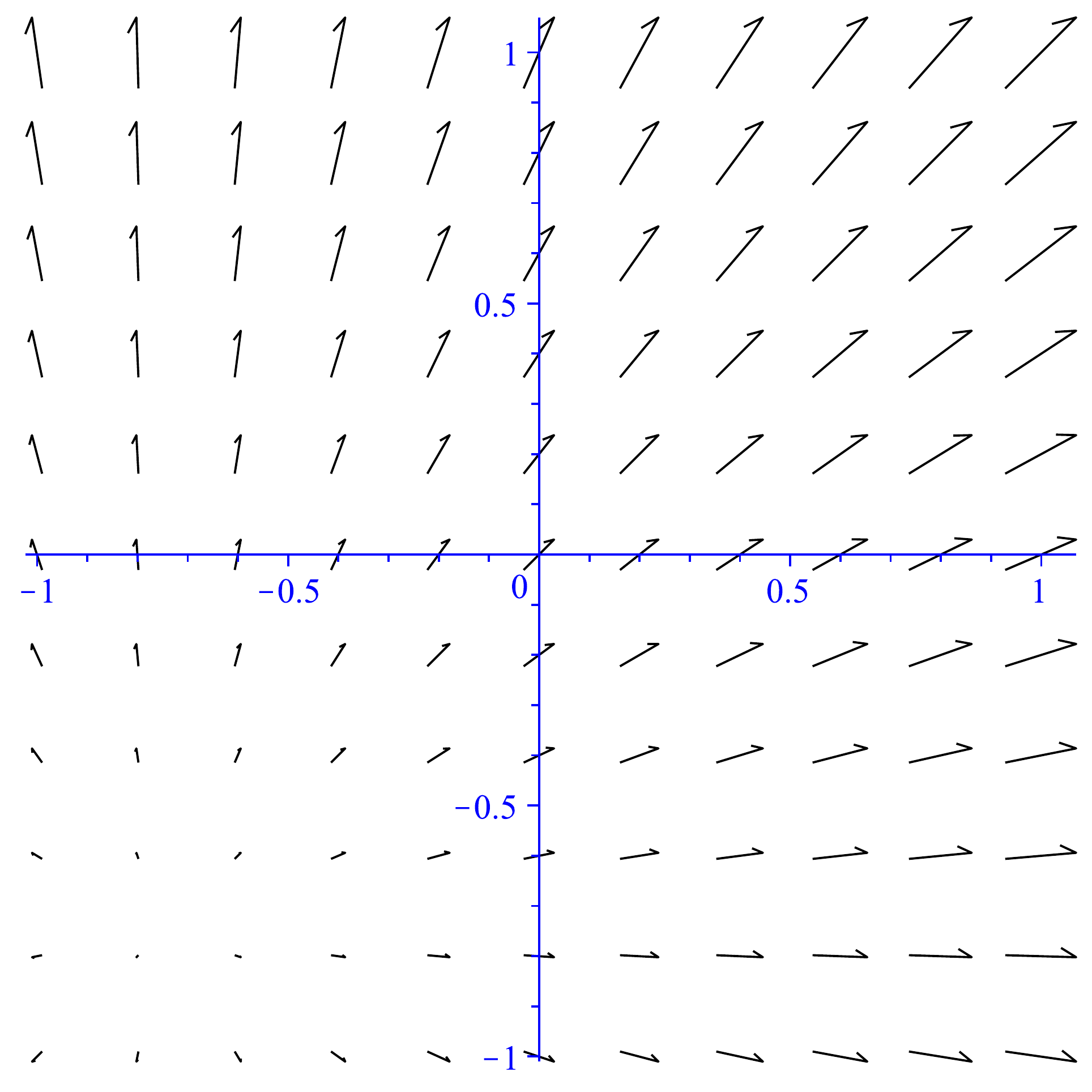} 
\\[10pt]
\includegraphics[width=5cm]{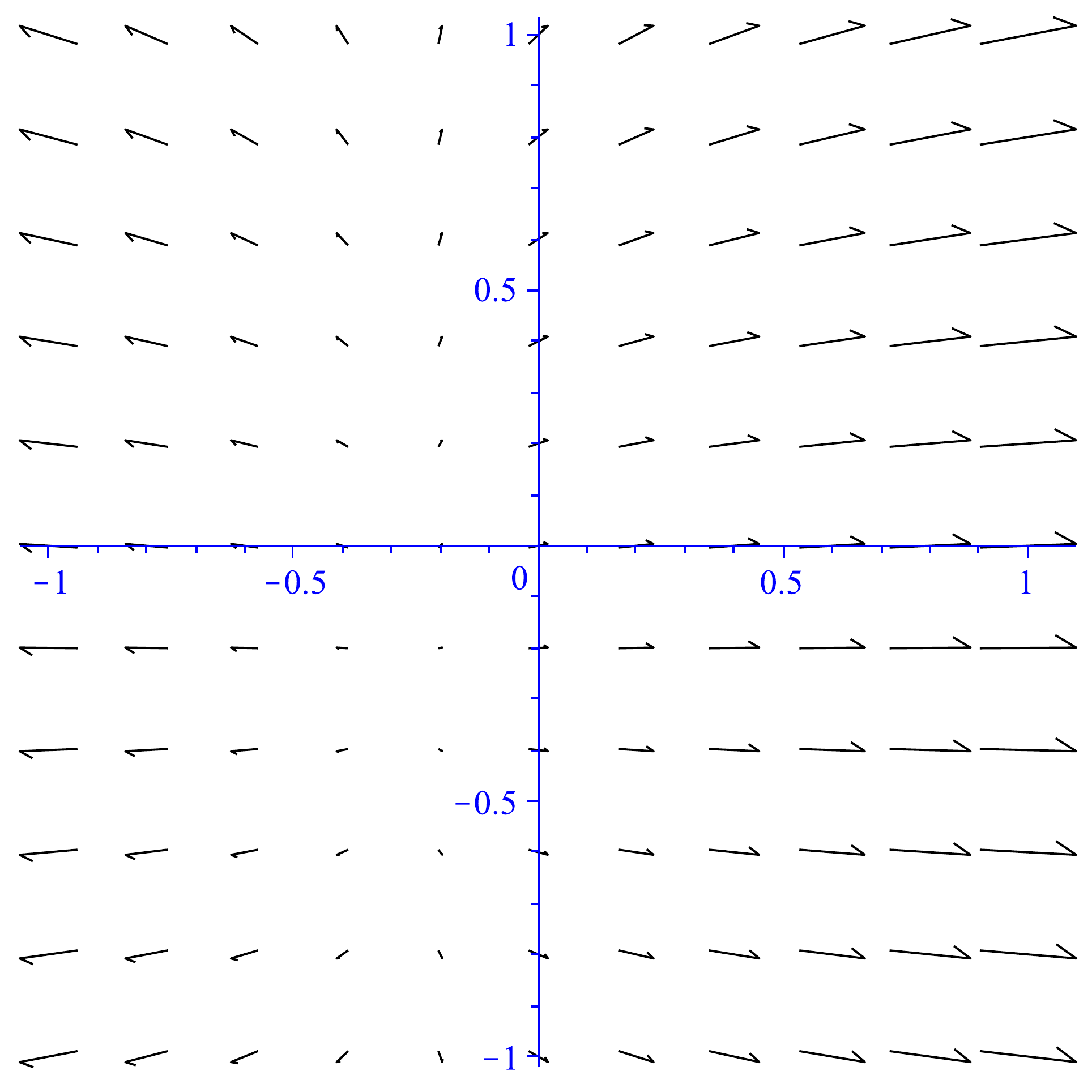}\qquad 
\includegraphics[width=5cm]{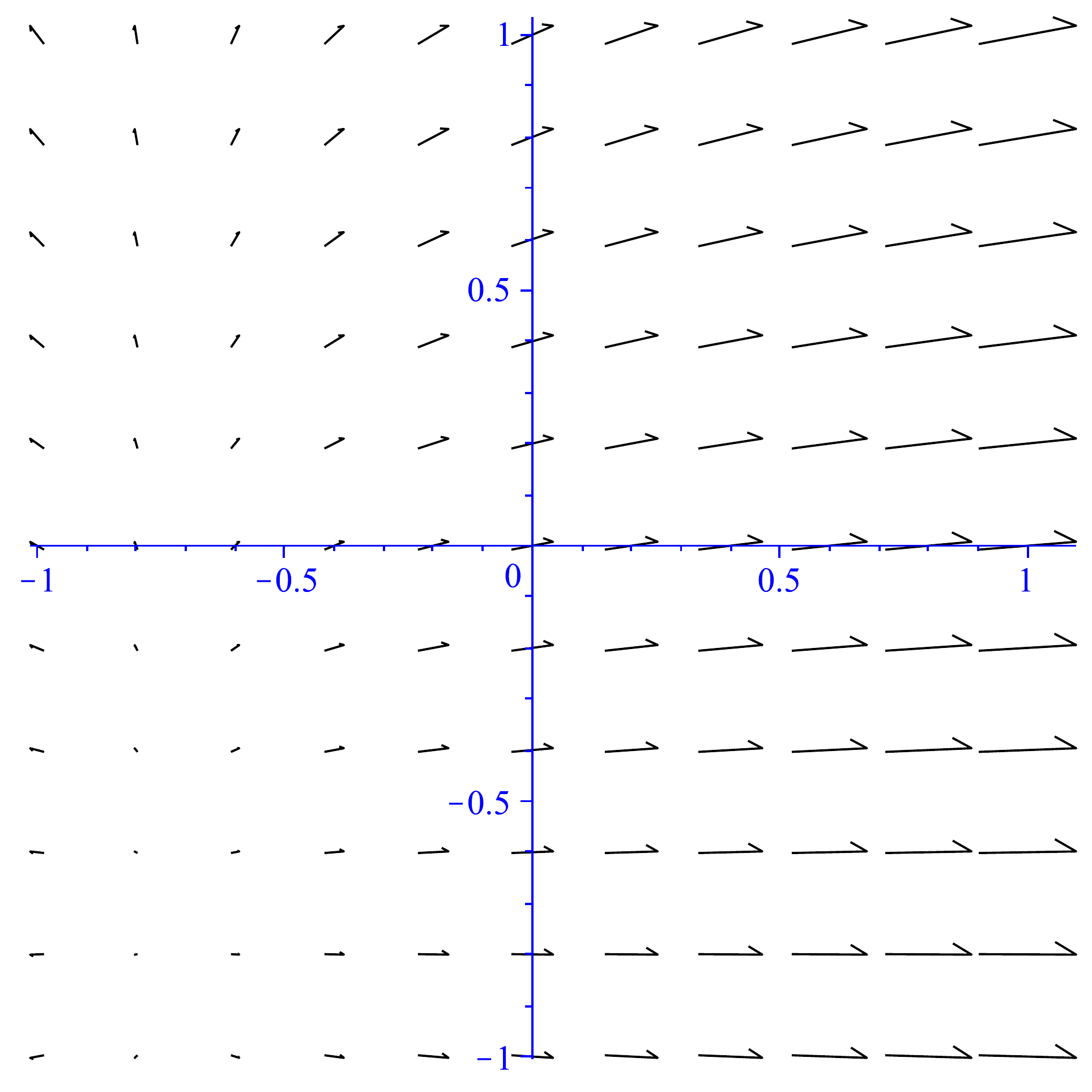} 
\caption{Projections of solution (\ref{eqx5x6r5}) $xy$-planes at $z=1/4$ (left) and at $z=3/4$ (right). Top: $a=b=1$, bottom: $a=1,b=10$}\label{fig:x5x6r5}
\end{figure}

A solution of (\ref{eqx5x6}) invariant under $R_5$ takes the form:
\begin{eqnarray}u_1(t,z)=u_2(t)z, \qquad v_1(t,z)=v_2(t)z, \qquad w_1(t,z)=w_2(t)z, \\[10pt] p_1(t,z)=p_2(t),\qquad \rho_1(t,z)=\rho_2(t)z^{-2}. \end{eqnarray}
Inserting these expressions into (\ref{eqx5x6}) gives:
\begin{eqnarray}  u_2(t)=\cfrac{u_3}{(t+a)(t+c)}, \qquad v_2(t)=\cfrac{v_3}{(t+b)(t+c)}, \qquad w_2(t)=\cfrac{1}{t+c}, \\[10pt]
\rho_2(t)=\cfrac{(t+c)\rho_3}{(t+a)(t+b)}\end{eqnarray}
where $u_3$, $v_3$ and $\rho_3$ are arbitrary constants. The pressure reads:
$$p_2(t)=\cfrac{\mu R\, h(t)^{-\frac{R}{C_v}-1}}{3C_v} h_c(t)f(t)\int \cfrac{h(t)^{\frac{R}{C_v}-1} g(t)}{h_c(t)f(t)} \d t$$
with
\begin{equation} h(t)=(t+a)(t+b)(t+c), \hspc h_c(t)=(t+c)^\frac{2\kappa(a-c)(b-c)}{\rho_3C_v}, \end{equation}
\begin{equation} f(t)=\exp\left({\frac{\kappa t(t+2a+2b-2c)}{\rho_3C_v}}\right) \end{equation}
and
\begin{eqnarray}
	g(t)=4(a^2+b^2+c^2-ab-bc-ca)t^2 \\[5pt]\phantom{g(t)}
	+4[a^2(b+c)+b^2(c+a)+c^2(a+b)-6abc]t \\[5pt]\phantom{g(t)}
	+4[a^2b(b-c)+b^2c(c-a)+c^2a(a-b)]\\[5pt]\phantom{g(t)}
+3[v_3^2(t+a)^2+u_3^2(t+b)^2].\end{eqnarray}
At last,
\be\begin{mybox}
	u(t,x,y,z)=\cfrac{u_3z+x(t+c)}{(t+a)(t+c)}, \qquad v(t,x,y,z)=\cfrac{v_3z+y(t+c)}{(t+b)(t+c)}, \\[10pt]
w(t,x,y,z)=\cfrac{z}{t+c}, 
\qquad p(t,x,y,z)=p_2(t), \\[10pt]  \rho(t,x,y,z)=\cfrac{\rho_3(t+c)}{(t+a)(t+b)z^2}.
\\[-10pt]\label{eqx5x6r5}\end{mybox}\ee
The projection of $\vt u$ on $xy$-planes are sketched in Figure \ref{fig:x5x6r5} for $c=0$, $u_3=v_3=1$ at $t=1$. The center-point, at which the velocity is vertical, is located at
\be x=\frac{-u_3z}{t+c}, \hspc y=\frac{-v_3z}{t+c}.
\label{eqx5x6r5_centerpoint}
\ee

\begin{figure}[ht]\centering
\includegraphics[width=5cm]{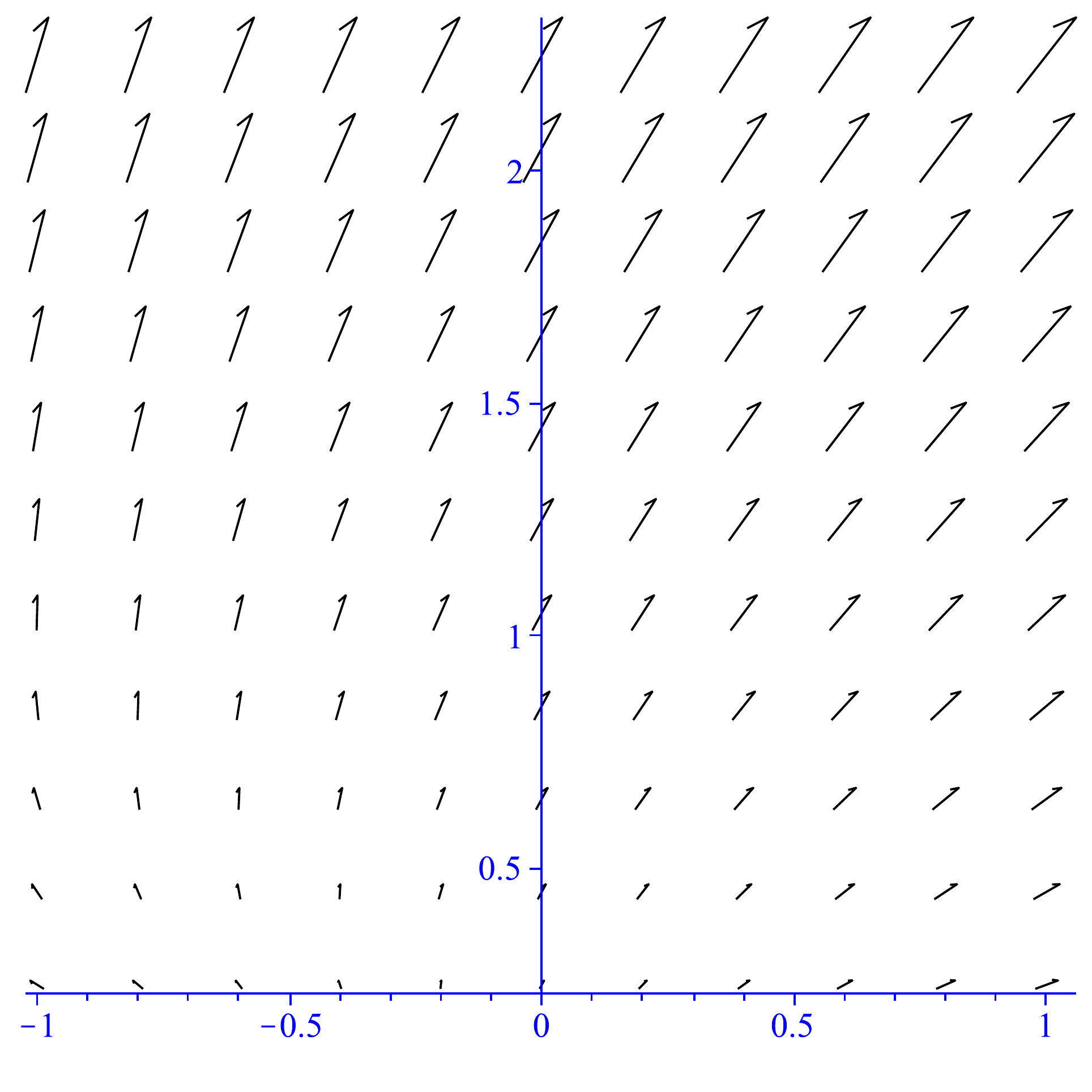} \qquad
\includegraphics[width=5cm]{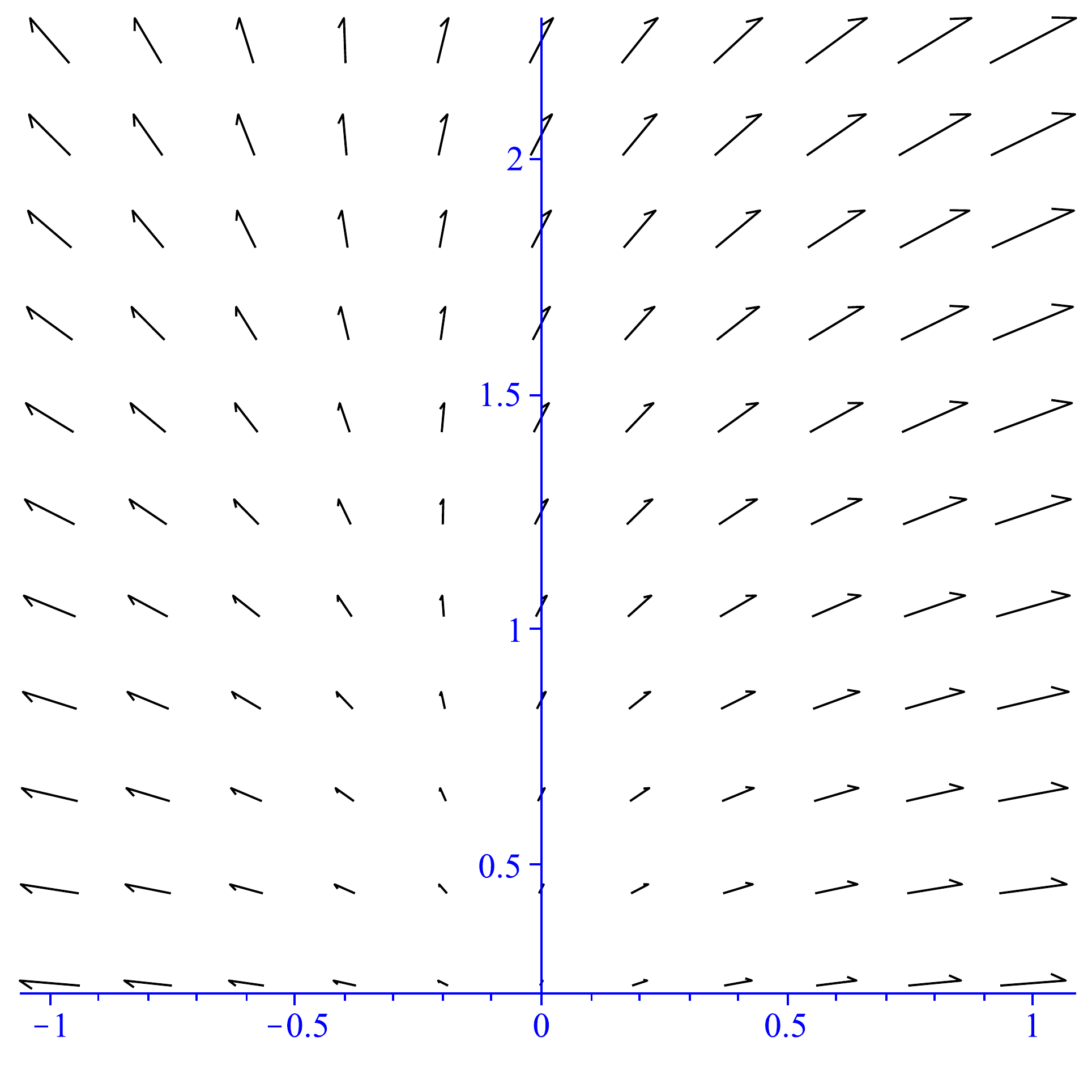} \\[10pt]
\includegraphics[width=5cm]{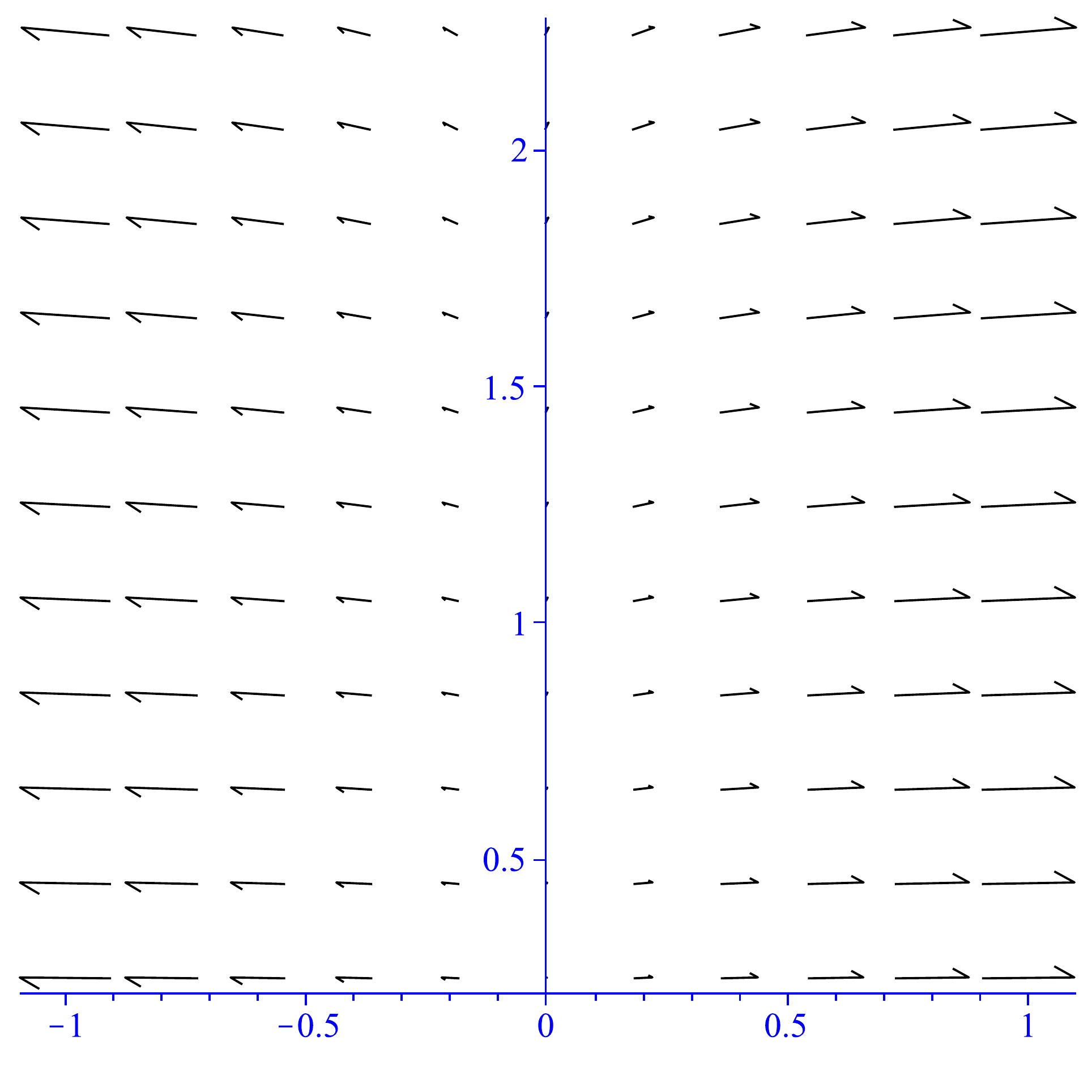} 
\caption{Projection of solution (\ref{eqx5x6r5}) on $xz$-plane when $a=b=1$. Top left: $c=0$, top right: $c=5$, bottom: $c=50$}\label{fig:x5x6r5_xz}
\end{figure}

As can be observed, the $xy$-plane-projected flow is radial around the center-point defined by equation (\ref{eqx5x6r5_centerpoint}) when $a=b=1$. It is not the case any longer if $a≠b$. The value of $c$ changes the position of the center-point (\ref{eqx5x6r5_centerpoint}) but not the shape of the $xy$-projections. It however has more influence on the projection of $\vt u$ on $xz$- or $yz$-plane. It can be stated in Figure \ref{fig:x5x6r5_xz}.

It is noteworthy that the generator
\begin{equation}\pd{}{z}+f(t)\pd{}{p} \label{y}\end{equation}
is an infinitesimal symmetry of the four first equations of (\ref{eqx5x6}) for any regular function $f$. It leads to the ansatz
\begin{eqnarray}
 u_1(t,z)=u_2(t),\qquad v_1(t,z)=v_2(t),\qquad w_1(t,z)=w_2(t),\\\\ p_1(t,z)=f(t)z+p_2(t),\qquad \rho_1(t,z)=\rho_2(t).
\end{eqnarray}
From the three first equations of (\ref{eqx5x6}), this ansatz allows to obtain:
\begin{eqnarray}
 u_2=\cfrac{u_3}{t+a}, \qquad v_2=\cfrac{v_3}{t+b}, \qquad \rho_2=\cfrac{\rho_3}{(t+a)(t+b)}, \\\\ f(t)=\cfrac{-\rho_3w_2'(t)}{(t+a)(t+b)}
\end{eqnarray}
where $u_3$, $v_3$ and $ρ_3$ are arbitrary constants
The last equation of (\ref{eqx5x6}) becomes:
\begin{eqnarray}
 -3\rho_3[C_vw_2''(t)(t+a)(t+b)+Rw'_2(t)(a+b+2t)]z+\Phi(t)
\end{eqnarray}
where $\Phi(t)$ is the $z$-independant rest of the equation. As seen, this equation still contains the variable $z$. 
It has not been reduced because vector field (\ref{y}) is not a symmetry of the last equation of (\ref{eqx5x6}). Cancelling the coefficient of $z$ and $\Phi(t)$, we get:
\begin{equation}
 w_2(t)=w_3\int h(t)^{-\frac{R}{C_v}}\d t \eqspc{with} h(t)=(t+a)(t+b),
\end{equation}
$w_3$ being a constant, and
\begin{equation}
 p_2(t)=h(t)^{-\frac{R}{C_v}-1}\int \left[\rho_3w_3w_2(t)+\cfrac{ 4R\mu  }{3C_v}\  g(t)
\ 
h(t)^{\frac{R}{C_v}-1}\right]\d t
\end{equation}
with 
\begin{equation}
 g(t)=(t+a)(t+b)+(a+b)^2.
\end{equation}
It follows that:
\be\begin{mybox}
	u(t,x,y,z)=\cfrac{x+u_3}{t+a}, \qquad v(t,x,y,z)=\cfrac{y+v_3}{t+b}, \\[10pt]
w(t,x,y,z)=w_3\displaystyle\int h(t)^{-\frac{R}{C_v}}\d t,  \qquad  \rho(t,x,y,z)=\cfrac{\rho_3}{h(t)},
\\[10pt] p(t,x,y,z)=p_2(t)-\rho_3w_3zh(t)^{-1-\frac{R}{C_v}}.
\\[-10pt]\end{mybox}\ee

Lastly, consider the symmetry generator $X_1-aX_2-bX_3-cX_4$ where $a$, $b$ and $c$ are constants. A basis of invariants under this vector field is
\[
ξ=t+ax+by+cz,\]\[ u_1=u,\qquad v_1=v,\qquad w_1=w,\qquad p_1=p,\qquad ρ_1=ρ.
\]
The equations can then be reduced with the ansatz
\[
u(t,x,y,z)=u_1(ξ),\qquad
v(t,x,y,z)=v_1(ξ),\qquad
w(t,x,y,z)=w_1(ξ),\]\[
p(t,x,y,z)=p_1(ξ),\qquad
ρ(t,x,y,z)=ρ_1(ξ).
\]
Indeed, system (\ref{nsc}) become
\begin{equation}
	\begin{cases}
		(ρ_1(1+D))'=0
		\\\displaystyle 
		ρ_1(1+D)u_1'+ap_1'-µ⦅\frac a3D+Au_1⦆''=0
		\\\displaystyle 
		ρ_1(1+D)v_1'+bp_1'-µ⦅\frac b3D+Av_1⦆''=0
		\\\displaystyle 
		ρ_1(1+D)w_1'+cp_1'-µ⦅\frac c3D+Aw_1⦆''=0
		\\\displaystyle 
		C_vp_1'(1+D)+(C_v+R)p_1D'+Rµ\tsr S_2-κA⦅÷{p_1}{ρ_1}⦆''=0
	\end{cases}
\end{equation}
where $A=a^2+b^2+c^2$, $D$ is the function of $ξ$ defined by
\[
D=au_1+bv_1+cw_1
\]
and
\[
\begin{array}{l}\displaystyle 
	\tsr S_2=÷23(D')^2+2(au'_1)^2+2(bv'_1)^2+2(cw'_1)^2
	\\[10pt]\phantom{\tsr S_2=}
	+(bu'_1+av'_1)^2+(cu'_1+bw'_1)^2+(cv'_1+bw'_1)^2.
\end{array}
\]
After integration, the first four equations of (\ref{eq_travel}) become
\begin{equation}
	\begin{array}{l}
		ρ_1(1+D)=ρ_2
		\\\displaystyle 
		ρ_2u_1+ap_1-µ⦅\frac a3D+Au_1⦆'=ρ_2u_2
		\\\displaystyle 
		ρ_2v_1+bp_1-µ⦅\frac b3D+Av_1⦆'=ρ_2v_2
		\\\displaystyle 
		ρ_2w_1+cp_1-µ⦅\frac c3D+Aw_1⦆'=ρ_2w_2
	\end{array}
	\label{eq_travel}
\end{equation}
for some constants $ρ_2$, $u_2$, $v_2$ and $w_2$. These equations can be used to express $p_1$, $v_1$ and $w_1$ as functions of $u_1$. Using the last equation of (\ref{eq_travel}) and going back to the original variables, we get a traveling wave solution of (\ref{nsc}):
\begin{equation}
	\begin{mybox}\\[-10pt]
		u=u_3\e^{÷{ρ_2ξ}{µA}}+u_4, 
		\\[10pt]\displaystyle 
		v=v_3\e^{÷{ρ_2ξ}{µA}}+v_2+b÷{u_4-u_2}a, \qquad w=w_3\e^{÷{ρ_2ξ}{µA}}+w_2+c÷{u_4-u_2}a,
		\\[10pt]\displaystyle 
		p=ρ_2÷{(au_3+bv_3+cw_3)\e^{÷{ρ_2ξ}{µA}}}{3A}+ρ_2÷{u_2-u_4}{a},
		\\[10pt]\displaystyle 
		ρ=÷{ρ_2}{(au_3+bv_3+cw_3)\e^{÷{ρ_2ξ}{µA}}+1+au_2+bv_2+cw_2 + A\cfrac{u_4-u_2}a }
	\end{mybox}
	\label{eq:travel}
\end{equation}
where 
\[ξ=t+ax+by+cz,\]
In (\ref{eq:travel}) $u_3$, $u_4$, $v_3$ and $w_3$ are constants linked by
\begin{equation}
	µAR(u_3^2+v_3^2+w_3^2)+(4κ-2µC_v)(au_3+bv_3+cw_3)^2=0
	\label{eq:travel_cond1}
\end{equation}
and either
\begin{equation}
	au_3+bv_3+cw_3=0
	\label{eq:travel_cond2}
\end{equation}
which leads to $u_3=v_3=w_3=0$ and corresponds to a constant solution or
\[
a(C_vµ-κ)(1+au_2+bv_2+cw_2)+A[(2C_v+3R)µ-2κ](u_2-u_4)=0.
\]
Disregarding condition (\ref{eq:travel_cond2}), it can be seen from condition (\ref{eq:travel_cond1}) that this traveling wave solution exists only when $2κ\leq µC_v$.
%
%

\section{Conclusion}

The infinitesimal Lie symmetries of the compressible Navier-Stokes equations were computed. The corresponding Lie group action were presented. From the commutation table, the Levi decomposition of the Lie algebra were presented.

Self-similar solutions were computed from the symmetries of the equations and successive reductions. These solutions represents many types of model flows. One can cite for example flows representing bidemensional vortices, evolving as $r^{-1}$ or $r^{-2}$ from a well or toward a sink point. Bidimensional solutions depending exponentially on $y$ could also be obtained. One can also cite the three dimensional vortex-like and traveling wave solutions.

Note that, since a Lie-symmetry takes a solution into another one, many other solutions from those presented here can be obtained, by composing them by transformations (\ref{time})--(\ref{scale2}).

In the present analysis, we did not intend to be exhaustive. Many other combinations of the presented infinitesimal generators may lead to completely new solutions.

\appendix
\section{Components of the equations and the infinitesimal generators\label{componentwise}}

The componentwise expression of equations (\ref{nsc}) is
\begin{equation}\begin{cases}
		ρ_t+uρ_x+vρ_y+wρ_z+\rho(u_x+v_y+w_z)=0 
\\[5pt]\displaystyle 
\rho⦅u_t+uu_x+vu_y+wu_z⦆= -p_x+µ⦅÷{4u_{xx}+v_{xy}+w_{xz}}3+u_{yy}+u_{zz}⦆
\\[10pt]\displaystyle 
\rho⦅v_t+uv_x+vv_y+wv_z⦆= -p_y+µ⦅÷{u_{xy}+4v_{yy}+w_{yz}}3+v_{xx}+v_{zz}⦆
\\[10pt]\displaystyle 
\rho⦅w_t+uw_x+vw_y+ww_z⦆= -p_z+µ⦅÷{u_{xz}+v_{yz}+4w_{zz}}3+w_{xx}+w_{yy}⦆
\\[10pt]\displaystyle 
\cfrac{C_v}{R}\left(p_t+up_x+vp_y+wp_z\right)=-⦅÷{C_v}R+1⦆p(u_x+v_y+w_z)
\\[5pt]\displaystyle 
\quad\quad-÷23µ(u_x+v_y+w_z)^2
\\[5pt]\displaystyle 
\quad\quad+µ⟦2⦅u_x^2+v_y^2+w_z^2⦆+⦅u_y+v_x⦆^2+⦅u_z+w_x⦆^2+⦅v_z+w_y⦆^2⟧
\\[5pt]\displaystyle 
\quad\quad+÷{κ}{R}⟦÷{p_{xx}+p_{yy}+p_{zz}}{ρ}-2÷{p_xρ_x+p_yρ_y+p_zρ_z}{ρ^2}\right.
\\\displaystyle
\quad\quad\quad\quad\left.-p÷{ρ_{xx}+ρ_{yy}+ρ_{zz}}{ρ^2}+2p÷{ρ_x^2+ρ_y^2+ρ_z^2}{ρ^3}⟧.
\end{cases}\label{nsc2}
\end{equation}

The first-order-derivative part $X^{(1)}$ of the prolonged vector field $pr^{(2)}X$ writes:
\begin{equation}
	\begin{array}{lcl}
		X^{(1)}&=&\displaystyle \xi^t_u\pdfrac{}{u_t}+\xi^x_u\pdfrac{}{u_x}+\xi^y_u\pdfrac{}{u_y}+\xi^z_u\pdfrac{}{u_z}\\[6pt]
		&+&\displaystyle\xi^t_v\pdfrac{}{v_t}+\xi^x_v\pdfrac{}{v_x}+\xi^y_v\pdfrac{}{v_y}+\xi^z_v\pdfrac{}{v_z}\\[6pt]
		&+&\displaystyle\xi^t_w\pdfrac{}{w_t}+\xi^x_w\pdfrac{}{w_x}+\xi^y_w\pdfrac{}{w_y}+\xi^z_w\pdfrac{}{w_z}\\[6pt]
		&+&\displaystyle\xi^t_p\pdfrac{}{p_t}+\xi^x_p\pdfrac{}{p_x}+\xi^y_p\pdfrac{}{p_y}+\xi^z_p\pdfrac{}{p_z}\\[6pt]
		&+&\displaystyle\xi^t_\rho\pdfrac{}{\rho_t}+\xi^x_\rho\pdfrac{}{\rho_x}+\xi^y_\rho\pdfrac{}{\rho_y}+\xi^z_\rho\pdfrac{}{\rho_z}.
	\end{array}
	\label{x1components}
\end{equation}
The componentes of $X^{(1)}$ can be expressed as functions of the $ξ^r$ and $ξ_q$ from relation (\ref{x1relation}). 
For example,
\begin{equation}
	\begin{array}{rcl}
		ξ^t_{ρ}&=&D_tξ_{ρ}-{ρ}_tD_tξ^t-{ρ}_xD_tξ^x-{ρ}_yD_tξ^y-{ρ}_zD_tξ^z\\\\
		&=&\quad\quad\pd {ξ_{ρ}}t+u_t\pd {ξ_{ρ}}u+v_t\pd {ξ_{ρ}}v+w_t\pd {ξ_{ρ}}w+p_t\pd {ξ_{ρ}}p+ρ_t\pd {ξ_{ρ}}{ρ}\\[10pt]
		&-&ρ_t⦅\pd {ξ^t}t\,+u_t\pd {ξ^t}u\,+v_t\pd {ξ^t}v\,+w_t\pd {ξ^t}w\,+p_t\pd {ξ^t}p\,+ρ_t\pd {ξ^t}{ρ}⦆\\[10pt]
		&-&ρ_x⦅\pd {ξ^x}t\,+u_t\pd {ξ^x}u\,+v_t\pd {ξ^x}v\,+w_t\pd {ξ^x}w\,+p_t\pd {ξ^x}p\,+ρ_t\pd {ξ^x}{ρ}⦆\\[10pt]
		&-&ρ_y⦅\pd {ξ^y}t\,+u_t\pd {ξ^y}u\,+v_t\pd {ξ^y}v\,+w_t\pd {ξ^y}w\,+p_t\pd {ξ^y}p\,+ρ_t\pd {ξ^y}{ρ}⦆\\[10pt]
		&-&ρ_z⦅\pd {ξ^z}t\,+u_t\pd {ξ^z}u\,+v_t\pd {ξ^z}v\,+w_t\pd {ξ^z}w\,+p_t\pd {ξ^z}p\,+ρ_t\pd {ξ^z}{ρ}⦆\\[10pt]
	\end{array}
\end{equation}
The second-order-derivative part $X^{(2)}$ can be written as a sum over the dependent variables $q=u,v,w,p,ρ$:
\[
	\begin{array}{l} 
		X^{(2)}=\displaystyle ∑_q\left(\xi^{tt}_q\pdfrac{}{q_{tt}}+\xi^{xx}_q\pdfrac{}{q_{xx}}+\xi^{yy}_q\pdfrac{}{q_{yy}}+\xi^{zz}_q\pdfrac{}{q_{zz}}+\xi^{tx}_q\pdfrac{}{q_{tx}}\right.
		\\[15pt]\displaystyle 
		\qquad\qquad\left.+ \xi^{ty}_q\pdfrac{}{q_{ty}}+\xi^{tz}_q\pdfrac{}{q_{tz}}
		+\xi^{xy}_q\pdfrac{}{q_{xy}}+\xi^{yz}_q\pdfrac{}{q_{yz}}+\xi^{zx}_q\pdfrac{}{q_{zx}}\right),
	\end{array}
\]
that is
\begin{equation}
	\begin{array}{lcl} 
		X^{(2)}&=&\displaystyle \xi^{tt}_u\pdfrac{}{u_{tt}}+\xi^{xx}_u\pdfrac{}{u_{xx}}+\xi^{yy}_u\pdfrac{}{u_{yy}}+\xi^{zz}_u\pdfrac{}{u_{zz}}
		\\[10pt]
		&&\displaystyle +\xi^{tx}_u\pdfrac{}{u_{tx}}+\xi^{ty}_u\pdfrac{}{u_{ty}}+\xi^{tz}_u\pdfrac{}{u_{tz}}
		\\[10pt]
		&&\displaystyle +\xi^{xy}_u\pdfrac{}{u_{xy}}+\xi^{yz}_u\pdfrac{}{u_{yz}}+\xi^{zx}_u\pdfrac{}{u_{zx}}
		\\[10pt]
		&+& \cdots \quad\text{(similar terms with $v$, $w$, $p$, and $ρ$ in place of $u$)}
			\end{array}
\end{equation}
Its coefficients can be computed from relation (\ref{x2relation}). 
An example of second order coefficient is:
\[ ξ^{xx}_u=D_xξ^x_u-u_{xt}D_xξ^t-u_{xx}D_xξ^x-u_{xy}D_xξ^y-u_{xz}D_xξ^z .\]
After developping the total derivatives, one finds the expression of $ξ^{xx}_u$ as a function the partial derivatives of the $ξ^r$ and $ξ_q$ up to second order.

To compute symmetries, $pr^{(2)}X$ is applied on equation (\ref{nsc2}). For instance, the action of $pr^{(2)}X$ on the first component of  equation (\ref{nsc2}) gives:
\[
\begin{array}{l}
	ξ^t_{ρ}+ξ_uρ_x+uξ^x_{ρ}+ξ_vρ_y+vξ^y_v+ξ_wρ_z+wξ^z_w
	\\[5pt]
	\qquad\qquad+ξ_{ρ}(u_x+v_y+w_z)+ρ(ξ^x_u+ξ^y_v+ξ^z_w)=0.
\end{array}
\]
It is a polynomial equation on the jet variables $t,\x,\u_{(2)},p_{(2)},ρ_{(2)}$, the coefficients being functions of the $ξ^r$ and $ξ_q$ and their derivatives.
The same procedure is applied to the four last components of equation (\ref{nsc2}) (and to their differential consequences, see \cite{olver86}).  The application of the infinitesimal symmetry condition~(\ref{symcond}) leads then to a system of polynomial equations on $u,v,w,p,ρ$, $u_t,u_x,\cdots$ (up to second order derivatives). Equating that the coefficients of these polynomials are zero, and with the help of a symbolic software \cite{cheviakov14}, one gets a system of partial differential equations on the $\xi^r$ and $ξ_q$, the solution of which is
\[
ξ^t=c_1t+2c_{11}t,\quad\quad 
ξ_p=-2c_{11}p,\quad\quad
ξ_{ρ}=-2c_{12}ρ\]
\[ξ^x=c_2x+c_5t+c_9z-c_{10}y+c_{11}x+c_{12}x\]
\[ξ^y=c_3y+c_6t-c_8z+c_{10}x+c_{11}y+c_{12}y\]
\[ξ^z=c_4z+c_7t+c_8y-c_9x+c_{11}z+c_{12}z \] 
\[ ξ_u=c_5+c_9w-c_{10}v-c_{11}u+c_{12}u\]
\[ ξ_v=c_6-c_8w+c_{10}u-c_{11}v+c_{12}v, \]
\[ ξ_w=c_7+c_8v-c_9u-c_{11}w+c_{12}w\]
In these expressions, the $c_i$ are arbitrary scalars. Therefore,
\[
X=∑_{i=1}^{12}c_iX_i
\]
where the $X_i$ are given in section \ref{symmetry}.


\begin{thebibliography}{10}

\bibitem{batchelor_2000}
G.K. Batchelor.
\newblock {\em An Introduction to Fluid Dynamics}.
\newblock Cambridge Mathematical Library. Cambridge University Press, 2000.

\bibitem{bluman02}
G.~Bluman and S.~Anco.
\newblock {\em Symmetry and Integration Methods for Differential Equations},
  volume 154 of {\em Applied Mathematical Sciences}.
\newblock Springer, 2002.

\bibitem{cheviakov14}
A.~Cheviakov.
\newblock Symbolic computation of nonlocal symmetries and nonlocal conservation
  laws of partial differential equations using the {GeM} package for {M}aple.
\newblock In J.-F. Ganghoffer and I.~Mladenov, editors, {\em Similarity and
  Symmetry Methods: Applications in Elasticity and Mechanics of Materials},
  volume~73 of {\em Lecture Notes in Applied and Computational Mechanics},
  pages 165--184. Springer International Publishing, 2014.

\bibitem{colonius91}
T.~Colonius, S.~Lele, and P.~Moin.
\newblock The free compressible viscous vortex.
\newblock {\em Journal of Fluid Mechanics}, 230:45--73.

\bibitem{curle71}
N~Curle and H.~Davies.
\newblock {\em Modern Fluid Dynamics. Volume 2: Compressible flow}.
\newblock Van Nostrand Reinhold Company, 1971.

\bibitem{fushchych94}
W.~Fushchych and R.~Popowych.
\newblock Symmetry reduction and exact solutions of the {N}avier-{S}tokes
  equations {I}, {II}.
\newblock {\em Journal of Nonlinear Mathematical Physics}, 1(1), 1994.

\bibitem{garnier09}
E.~Garnier, N.~Adams, and P.~Sagaut.
\newblock {\em Large eddy simulation for compressible flows}.
\newblock Scientific Computation. Springer, 2006.

\bibitem{grassi00}
V.~Grassi, R.A. Leo, G.~Soliani, and P.~Tempesta.
\newblock Vorticies and invariant surfaces generated by symmetries for the 3{D}
  {N}avier-{S}tokes equation.
\newblock {\em Physica A}, 286:79--108, 2000.

\bibitem{hydon00a}
P.E. Hydon.
\newblock {\em Symmetry methods for differential equations. {A} beginner's
  guide.}
\newblock Cambridge University Press, 2000.

\bibitem{ibragimov}
N.H. Ibragimov.
\newblock {\em CRC handbook of {L}ie group analysis of differential equations.
  Vol 1-3}.
\newblock CRC Press, 1996.

\bibitem{khujadze04}
G.~Khujadze and M.~Oberlack.
\newblock {DNS} and scaling laws from new symmetry groups of {ZPG} turbulent
  boundary layer flow.
\newblock {\em Theoretical and Computational Fluid Dynamics}, 18(5):391--411,
  2004.

\bibitem{oberlack01}
M.~Oberlack.
\newblock A unified approach for symmetries in plane parallel turbulent shear
  flows.
\newblock {\em Journal of Fluid Mechanics}, 427:299--328, 2001.

\bibitem{olver10}
Olver, Lozier, Boisvert, and Clark.
\newblock {\em NIST Handbook of Mathematical Functions}.
\newblock Cambridge University Press, 2010.

\bibitem{olver86}
P.~Olver.
\newblock {\em Applications of {L}ie groups to differential equations}, volume
  107 of {\em Graduate texts in mathematics}.
\newblock Springer-{V}erlag, 1986.

\bibitem{nova09}
D.~Razafindralandy, A.~Hamdouni, and M.~Chhay.
\newblock {\em Numerical Simulation Research Progress}, chapter Symmetry in
  Turbulence Simulation, pages 161--207.
\newblock Nova Publishers, 2009.

\bibitem{ejm07}
D.~Razafindralandy, A.~Hamdouni, and M.~Oberlack.
\newblock Analysis and development of subgrid turbulence models preserving the
  symmetry properties of the {Navier--Stokes} equations.
\newblock {\em European Journal of Mechanics/B}, 26:531--550, 2007.

\bibitem{sachdev05}
P.~Sachdev, K.~Joseph, and M.~Haque.
\newblock Exact solutions of compressible flow equations with spherical
  symmetry.
\newblock {\em Studies in Applied Mathematics}, 114:325--342, 2005.

\bibitem{toutant17}
A.~Toutant.
\newblock General and exact pressure evolution equation.
\newblock {\em Physics Letters A}, 381(44):3739--3742, 2017.

\bibitem{tsangaris00}
S~Tsangaris and T.~Pappou.
\newblock Analytical solutions for the unsteady compressible flow equations
  serving as test cases for the verification of numerical schemes.
\newblock Technical report, Defense Technical Information Center, 2000.

\bibitem{unal97}
G.~{\"U}nal.
\newblock Constitutive equation of turbulence and the {L}ie symmetries of
  {N}avier-{S}tokes equations.
\newblock In N.H. Ibragimov, K.~Razi~Naqvi, and E.~Straume, editors, {\em
  Modern Group Analysis VII}, pages 317--323. Mars Publishers, 1997.

\bibitem{zhang92}
T.A. Zang, R.B. Dahlburg, and J.P. Dahlburg.
\newblock Direct and large‐eddy simulations of three‐dimensional
  compressible {N}avier–{S}tokes turbulence.
\newblock {\em Physics of Fluids A: Fluid Dynamics}, 4(1):127--140, 1992.

\end{thebibliography}

\end{document}